\input harvmac
\def\journal#1&#2(#3){\unskip, \sl #1\ \bf #2 \rm(19#3) }
\def\andjournal#1&#2(#3){\sl #1~\bf #2 \rm (19#3) }

\def\frac#1#2{{#1\over#2}}

\def\half{\frac12}

\def\d{\partial}

\def\inbar{\,\vrule height1.5ex width.4pt depth0pt}
\def\IC{\relax\hbox{$\inbar\kern-.3em{\rm C}$}}
\def\IR{\relax{\rm I\kern-.18em R}}
\def\IP{\relax{\rm I\kern-.18em P}}
\def\IZ{\relax{\rm I\kern-.18em Z}}

%
%

%
\catcode`\@=11
\def\slash#1{\mathord{\mathpalette\c@ncel{#1}}}
\overfullrule=0pt

\def\CC{{\cal C}}
\def\DD{{\cal D}}

\def\HH{{\cal H}}

\def\MM{{\cal M}}
\def\NN{{\cal N}}

\def\WW{{\cal W}}
\def\XX{{\cal X}}
\def\YY{{\cal Y}}
\def\ZZ{{\cal Z}}

\def\underrel#1\over#2{\mathrel{\mathop{\kern\z@#1}\limits_{#2}}}

\catcode`\@=12


%


\def\gs{{\bf G}}

\rightline{RI-3-00}
\Title{
\rightline{hep-th/0009242}}
{\vbox{\centerline{Superstrings on $AdS_3$ and Symmetric Products}}}
\medskip
\centerline{\it Riccardo Argurio, Amit Giveon and Assaf Shomer}
\bigskip
\centerline{Racah Institute of Physics, The Hebrew University}
\centerline{Jerusalem 91904, Israel}
\bigskip\bigskip\bigskip
\noindent
Some properties of $N=2$ superstrings on $AdS_3\times \NN$ are studied.
Their spectrum has the same pattern as 2-$d$ CFTs in the moduli space of
symmetric products $\MM^p/S_p$, where $p$ is associated with
the number of long fundamental strings required to construct the $AdS_3$ 
background, and $\MM$ is the spacetime CFT corresponding to short string 
excitations of a single long string.
Worldsheet operators associated with $w$ long strings 
correspond to twisted sectors of $\MM^p/S_p$.

\vfill

\Date{9/00}

\lref\fms{D.\ Friedan, E.\ Martinec, and S.\ Shenker, ``Conformal
invariance, supersymmetry, and string theory," Nucl.\ Phys.\ {\bf B271}
(1986) 93.}
\lref\gk{A. Giveon and D. Kutasov, ``Little string theory in a double scaling
limit,'' JHEP {\bf 9910} (1999) 034, hep-th/9909110;
``Comments on double scaled little string theory,'' 
JHEP {\bf 0001} (2000) 023, hep-th/9911039.}
\lref\ags{R.~Argurio, A.~Giveon and A.~Shomer,
``Superstring theory on $AdS_3\times G/H$ and boundary $N = 3$ superconformal
symmetry,'' JHEP {\bf 0004} (2000) 010, hep-th/0002104.}%
\lref\gks{A.~Giveon, D.~Kutasov and N.~Seiberg,
``Comments on string theory on $AdS_3$,''
Adv.\ Theor.\ Math.\ Phys.\  {\bf 2} (1998) 733,
hep-th/9806194.}%
\lref\ks{D.~Kutasov and N.~Seiberg,
``More comments on string theory on $AdS_3$,''
JHEP {\bf 9904} (1999) 008,
hep-th/9903219.}%
\lref\efgt{S.~Elitzur, O.~Feinerman, A.~Giveon and D.~Tsabar,
``String theory on $AdS_3 \times S^3 \times S^3 \times S^1$,''
Phys.\ Lett.\  {\bf B449} (1999) 180,
hep-th/9811245.}%
\lref\ss{A.~Schwimmer and N.~Seiberg,
``Comments on the $N=2, N=3, N=4$ superconformal algebras in two-dimensions,''
Phys.\ Lett.\  {\bf B184} (1987) 191.}%
\lref\dmvv{R.~Dijkgraaf, G.~Moore, E.~Verlinde and H.~Verlinde,
``Elliptic genera of symmetric products and second quantized strings,''
Commun.\ Math.\ Phys.\  {\bf 185} (1997) 197,
hep-th/9608096.}%
\lref\vw{C.~Vafa and E.~Witten,
``A strong coupling test of S duality,''
Nucl.\ Phys.\  {\bf B431} (1994) 3,
hep-th/9408074.}%
\lref\yis{S.~Yamaguchi, Y.~Ishimoto and K.~Sugiyama,
``$AdS_3$/CFT$_2$ correspondence and space-time N = 3 superconformal
algebra,'' JHEP {\bf 9902} (1999) 026,
hep-th/9902079.}%
\lref\fks{J.~Fuchs, A.~Klemm and M.~G.~Schmidt,
``Orbifolds by cyclic permutations in Gepner type superstrings and in
the corresponding Calabi-Yau manifolds,''
Annals Phys.\  {\bf 214} (1992) 221.}%
\lref\klsc{A.~Klemm and M.~G.~Schmidt,
``Orbifolds by cyclic permutations of tensor product conformal field
theories,'' Phys.\ Lett.\  {\bf B245} (1990) 53.}%
\lref\orbif{L.~Dixon, J.~A.~Harvey, C.~Vafa and E.~Witten,
``Strings on orbifolds,''
Nucl.\ Phys.\  {\bf B261} (1985) 678; L.~Dixon, D.~Friedan, E.~Martinec
and S.~Shenker,
``The conformal field theory of orbifolds,''
Nucl.\ Phys.\  {\bf B282} (1987) 13.}%
\lref\mardav{F.~Larsen and E.~Martinec,
``U(1) charges and moduli in the D1-D5 system,''
JHEP {\bf 9906} (1999) 019
hep-th/9905064, J.~R.~David,
``String theory and black holes,''
hep-th/9911003.}%
\lref\ms{J.~Maldacena and A.~Strominger,
``$AdS_3$ black holes and a stringy exclusion principle,''
JHEP {\bf 9812} (1998) 005,
hep-th/9804085.}%
\lref\dijk{R.~Dijkgraaf,
``Instanton strings and hyperKaehler geometry,''
Nucl.\ Phys.\  {\bf B543} (1999) 545,
hep-th/9810210.}
\lref\deboer{J.~de Boer,
``Six-dimensional supergravity on $S^3 \times AdS_3$ and $2d$ conformal field
theory,'' Nucl.\ Phys.\  {\bf B548} (1999) 139,
hep-th/9806104.}%
\lref\kll{D.~Kutasov, F.~Larsen and R.~G.~Leigh,
``String theory in magnetic monopole backgrounds,''
Nucl.\ Phys.\  {\bf B550} (1999) 183,
hep-th/9812027.}%
\lref\dBps{J.~de Boer, A.~Pasquinucci and K.~Skenderis,
``AdS/CFT dualities involving large $2d$ $N = 4$ superconformal symmetry,''
hep-th/9904073.}%
\lref\dst{F.~Defever, S.~Schrans and K.~Thielmans,
``Moding Of Superconformal Algebras,''
Phys.\ Lett.\  {\bf B212} (1988) 467. }%
\lref\sw{N.~Seiberg and E.~Witten,
``The D1/D5 system and singular CFT,''
JHEP {\bf 9904} (1999) 017,
hep-th/9903224.}%
\lref\mo{J.~Maldacena and H.~Ooguri,
``Strings in $AdS_3$ and $SL(2,R)$ WZW model. I,''
hep-th/0001053.}%
\lref\gr{A.~Giveon and M.~Rocek,
``Supersymmetric string vacua on $AdS_3\times \NN$,''
JHEP {\bf 9904} (1999) 019,
hep-th/9904024.}%
\lref\bl{D.~Berenstein and R.~G.~Leigh,
``Spacetime supersymmetry in $AdS_3$ backgrounds,''
Phys.\ Lett.\  {\bf B458} (1999) 297,
hep-th/9904040.}%
\lref\gkp{A.~Giveon, D.~Kutasov and O.~Pelc,
``Holography for non-critical superstrings,''
JHEP {\bf 9910} (1999) 035,
hep-th/9907178.}%
\lref\ofer{O.~Feinerman,
``String theory on $AdS_3\times S^3\times S^3\times S^1$,''
M.Sc Thesis (The Hebrew University of Jerusalem).}%
\lref\malpri{J.~Maldacena, private communication.}
\lref\difr{P.~Di Francesco, P.~Mathieu, and D.~S\'en\'echal
``Conformal Field Theory'' Springer (1997).}%
\lref\spec{R.~Argurio, A.~Giveon and A.~Shomer,
``Superstring theory on $AdS_3$ times a coset manifold,"
Proceedings of the conference Non-perturbative Quantum Effects 2000,
Paris, PRHEP-tmr2000/007, and paper to appear.}
\lref\sv{A.~Strominger and C.~Vafa,
``Microscopic Origin of the Bekenstein-Hawking Entropy,''
Phys.\ Lett.\  {\bf B379} (1996) 99,
hep-th/9601029.}
\lref\mmoores{J.~Maldacena, G.~Moore and A.~Strominger,
``Counting BPS black holes in toroidal type II string theory,''
hep-th/9903163.}

\newsec{Introduction and Summary}

\lref\early{A. B. Zamolodchikov and V. A. Fateev, Sov. J. Nucl. Phys. {\bf 43}
(1986) 657;
J. Balog, L. O'Raifeartaigh, P. Forgacs, and A. Wipf,
Nucl. Phys. {\bf B325} (1989) 225;
L. J. Dixon, M. E. Peskin and J. Lykken,  Nucl. Phys. {\bf B325} (1989)
329;
A. Alekseev and S. Shatashvili, Nucl. Phys. {\bf B323} (1989) 719;
N. Mohameddi, Int. J. Mod. Phys. {\bf A5} (1990) 3201;
P. M. S. Petropoulos, Phys. Lett. {\bf B236} (1990) 151;
M. Henningson and S. Hwang, Phys. Lett. {\bf B258} (1991) 341;
M. Henningson, S. Hwang, P. Roberts, and B. Sundborg, Phys. Lett. {\bf
B267} (1991) 350;
S. Hwang, Phys. Lett. {\bf B276} (1992) 451, hep-th/9110039;
I. Bars and D. Nemeschansky, Nucl. Phys. {\bf B348} (1991) 89;
S. Hwang, Nucl. Phys. {\bf B354} (1991) 100;
K. Gawedzki, hep-th/9110076;
I. Bars, Phys. Rev. {\bf D53} (1996) 3308, hep-th/9503205;
in {\it Future Perspectives In String Theory} (Los Angeles, 1995),
hep-th/9511187;
O. Andreev, Phys.Lett. {\bf B375} (1996) 60, hep-th/9601026,
Nucl.\ Phys.\  {\bf B552} (1999) 169, hep-th/9901118,
Nucl.\ Phys.\  {\bf B561} (1999) 413, hep-th/9905002, 
Phys.Rev. {\bf D61} (2000) 126001, hep-th/9909222;
J. L. Petersen, J. Rasmussen and M. Yu, hep-th/9607129, Nucl.Phys. {\bf
B481} (1996) 577;
Y. Satoh, Nucl. Phys. {\bf B513} (1998) 213, hep-th/9705208;
J. Teschner, Nucl.Phys. {\bf B546} (1999) 390, hep-th/9712256,  
Nucl.Phys. {\bf B546} (1999) 369, hep-th/9712258, 
Nucl.Phys. {\bf B571} (2000) 555, hep-th/9906215;
J.~M.~Evans, M.~R.~Gaberdiel and M.~J.~Perry,
Nucl.\ Phys.\  {\bf B535} (1998) 152, hep-th/9806024;
I.~Pesando, JHEP {\bf 9902} (1999) 007, hep-th/9809145, 
Mod.\ Phys.\ Lett.\  {\bf A14} (1999) 2561, hep-th/9903086;
J.~Rahmfeld and A.~Rajaraman, Phys.\ Rev.\  {\bf D60} (1999) 064014
hep-th/9809164;
S.~Mukherji and S.~Panda, Phys.\ Lett.\  {\bf B451} (1999) 53, hep-th/9810140;
K.~Ito, Phys.\ Lett.\  {\bf B449} (1999) 48, hep-th/9811002,
Mod.\ Phys.\ Lett.\  {\bf A14} (1999) 2379, hep-th/9910047;
J.~de Boer, H.~Ooguri, H.~Robins and J.~Tannenhauser,
JHEP {\bf 9812} (1998) 026, hep-th/9812046;
K.~Hosomichi and Y.~Sugawara, JHEP {\bf 9901} (1999) 013, hep-th/9812100,
JHEP {\bf 9907} (1999) 027, hep-th/9905004;
M.~Yu and B.~Zhang, Nucl.\ Phys.\  {\bf B551} (1999) 425, hep-th/9812216;
N.~Berkovits, C.~Vafa and E.~Witten, JHEP {\bf 9903} (1999) 018,
hep-th/9902098;
M.~Banados and A.~Ritz, Phys.\ Rev.\  {\bf D60}, 126004 (1999), hep-th/9906191;
I.~Bars, C.~Deliduman and D.~Minic, hep-th/9907087;
P.~M.~Petropoulos, hep-th/9908189;
Y.~Sugawara, Nucl.Phys. {\bf B576} (2000) 265, hep-th/9909146;
G.~Giribet and C.~Nunez, JHEP {\bf 9911} (1999) 031, hep-th/9909149;
A.~Kato and Y.~Satoh, Phys.Lett. {\bf B486} (2000) 306, hep-th/0001063;
M.~Langham, Nucl.Phys. {\bf B580} (2000) 565, hep-th/0002032;
J.~Rasmussen, Nucl.\ Phys.\  {\bf B582} (2000) 649, hep-th/0002188, 
hep-th/0003035; 
M.~O'Loughlin, gr-qc/0002092;
N.~Ishibashi, K.~Okuyama and Y.~Satoh, hep-th/0005152;
J.~Maldacena, H.~Ooguri and J.~Son, hep-th/0005183;
O.~Lunin and S.~D.~Mathur, hep-th/0006196;
A.~Nichols and Sanjay, hep-th/0007007;
A.~Ali, hep-th/0007021;
I.~Benkaddour, A.~El Rhalami and E.~H.~Saidi, hep-th/0007142;
A.~Lewis, hep-th/0009096;
J.~Jing and M.~Yu, hep-th/0009118.}
\lref\mms{J. Maldacena, J. Michelson and A. Strominger,
``Anti-de Sitter fragmentation," JHEP {\bf 9902} (1999) 011, hep-th/9812073.}
\lref\hhs{Y. Hikida, K. Hosomichi and Y. Sugawara,
``String theory on $AdS_3$ as discrete light-cone Liouville theory,"
hep-th/0005065.}

In this paper we continue the study of the superstring on $AdS_3\times \NN$.
We use the NSR formulation; for earlier works see, e.g., \refs{\early,
\gks,\efgt,\kll,\yis,\ks,\sw,\gr,\bl,\gkp,\mo,\ags,\hhs}.
The study in this work leads us to conjecture that the spacetime
theory is a two dimensional CFT in the moduli space of the symmetric
product:
\eqn\mkp{(\MM_{6k})^p/S_p~.}
Here $k$ is the level of the $AdS_3\simeq SL(2)$ worldsheet CFT.
$\MM_{6k}$ is a two dimensional CFT with a central charge $c=6k$,
the precise structure of which depends on the details of the internal
worldsheet CFT $\NN$.
The integer number $p$, which parametrizes the string coupling $g_s$,
is associated with the number
of fundamental ``long" strings required to construct the $AdS_3$ background.
More precisely, it has one of the following
(related) interpretations:
\item{(a)}
To leading order in $1/p$ \gks:
\eqn\gs{g_s^2={V_{\NN}\over 4p\sqrt{k}}~,}
where $V_{\NN}$ is the volume of the compactification manifold $\NN$
(measured in string units).
\item{(b)}
$p$ is the number of fundamental strings used to generate the
$AdS_3$ background \gkp:
starting with the background $R^{1,1}\times R_\phi \times \NN$,
where $R_\phi$ is the real line with a linear dilaton, one obtains the
$AdS_3\times \NN$ background by placing $p$ fundamental strings stretched
in $R^{1,1}$ and then going to their near horizon limit.
\item{(c)}
$p$ is the maximal number of ``long strings" \refs{\gks,\mms,\ks,\sw,\mo}
which can be emitted in the $AdS_3$ background and stretch in its boundary.
\item{(d)} The ``exclusion principle" \ms\ allows at most 
``$p$-particle'' (BPS) states.

\noindent
The spacetime Hilbert space of $\MM_{6k}$
consists of the ``single particle'' states
corresponding to worldsheet vertex operators in the ``short
string" sector, and the continuum of a single ``long string."

{}For concreteness of the discussion, we shall consider the family of
supersymmetric string backgrounds which lead to at least $N=2$
supersymmetry in spacetime \refs{\gr,\bl}. In particular,
we will verify that the pattern of the 
spacetime chiral spectrum of these theories is in
agreement with the picture conjectured above.

To do that,
we begin in section 2 by reviewing the basic properties of a bosonic string
propagating in $AdS_3$, and introduce a ``twist'' required to impose single
valuedness of the wave functions on $AdS_3$ ``intrinsically'' in the
$AdS_3$ worldsheet theory. This leads to the results of \mo, which were
obtained using spectral flow arguments. In particular, the sector of $w$
long strings in \mo\ is associated with the $w$'th twisted sector.

In section 3, we review the $N=2$ superstring on $AdS_3\times\NN$
and introduce the ``twist'' leading to the analogue of \mo\ in the case of
a superstring. In section 4, we study observables leading to chiral
``single particle states" of the spacetime $N=2$ SCFT.
In section 5, we show that the pattern of this chiral spectrum
coincides with the one of \mkp.
In particular, the chiral operators associated to $N$
long strings correspond to operators in the $Z_N$
twisted sector of $(M_{6k})^p/S_p$.
In section 6, we present some examples, and in section 7 we discuss a few
points. Finally, some of the technical details appear in the appendices.

A related work on string theory on $AdS_3\times \NN$ using
discrete light-cone Liouville theory
methods appeared in \hhs.

\newsec{The Bosonic String on $AdS_3$}

In this section we review some basic facts about bosonic
string propagation on $AdS_3$ space \refs{\early,\gks,\ks}. We then show how
properly taking into account single valuedness of the wave
function forces one to include new sets of operators which
were argued to exist in \mo, where reasoning based on spectral flow was used.

\subsec{Review}
String theory on an $AdS_3\times \NN$ background is
formulated on the worldsheet as the product of an $SL(2)$ WZW model
and the CFT on $\NN$.
$AdS_3\simeq SL(2)$ has three $SL(2)$ affine currents $J^a(z)$
and anti-currents $\bar J^a(\bar z)$, $a=\pm,3$,
generating an affine $SL(2)_L\times SL(2)_R$ algebra at level $k$.
{}For simplicity of the discussion we will many times refer
only to left-movers. As usual in string theory, the spacetime
theory has three conserved charges:
\eqn\sltwo{L_0=-\oint dz J^3(z)~, \qquad
L_1=-\oint dz J^+(z)~, \qquad L_{-1}=-\oint dz J^-(z)~,}
which satisfy the $SL(2)$ Lie algebra
$[L_n,L_m]=(n-m)L_{n+m}$, $n,m=0,\pm 1$.
A special property of string theory on $AdS_3$ is that
this Lie algebra is enhanced to the full Virasoro algebra:
\eqn\vir{[L_n,L_m]=(n-m)L_{n+m}+{c_{st}\over 12}(n^3-n)\delta_{n+m,0}~,
\qquad n\in {\bf Z}~.}
The construction of the worldsheet vertex operators
corresponding to the spacetime generators $L_n$ is given
in \refs{\gks,\ks}. For future purpose, we pick up the
presentation of $L_n$ in the free field realization of $SL(2)$:
\eqn\ln{L_n=\oint dz \left[(n^2-1)J^3\gamma^n
-{n(n-1)\over 2}J^-\gamma^{n+1}
-{n(n+1)\over 2}J^+\gamma^{n-1}\right]~.}
Here $\gamma$ is the dimension zero field in the free field
representation of $SL(2)$; for the details of the construction
\ln\ see \gks.
The spacetime central charge $c_{st}$ is related to the
worldsheet $SL(2)$ level $k$ via:
\eqn\cst{c_{st}=6kp~,}
where $p$ is the integer number discussed in the introduction \gs.

Let us introduce a canonically normalized scalar
$X$: $X(z)X(z')\sim -\log(z-z')$,
in terms of which we write:
\eqn\jthree{J^3=-\sqrt{k\over 2}\d X~.}
We can now decompose any operator $\Phi$ in $AdS_3$
to $SL(2)/U(1)\times U(1)_X$, where the $U(1)_X$ is the one generated
by $J^3$; the ``parafermionic'' $SL(2)/U(1)$ part of $\Phi$
will be denoted by $\Psi$.
In particular, the charged currents decompose as:
\eqn\jpm{J^{\pm}=\Psi_{\pm}e^{\pm\sqrt{2\over k}X}~.}
The bosonic string on $AdS_3\times \NN$ has physical vertex operators
$V(j;m,\bar{m})$
which are labeled, in particular, by their $SL(2)$ numbers $j,m,\bar{m}$,
where $j$ is related to the second Casimir of $SL(2)$, $-j(j+1)$, and
$m, \bar m$ are the eigenvalues of $J^3, \bar J^3$, respectively.
$V(j;m,\bar{m})$ includes the contribution of a function $\Phi_{jm\bar{m}}$
on $AdS_3$ which decomposes as:
\eqn\phipsi{\Phi_{jm\bar{m}}=\Psi_{jm\bar{m}}
e^{\sqrt{2\over k}(mX(z)+\bar m\bar X(\bar z))}~,}
where the $SL(2)/U(1)$ part has left-handed scaling dimension
\eqn\dimpsi{\Delta(\Psi_{jm\bar{m}})={-j(j+1)\over k-2}+{m^2\over k}~,}
and similarly for the right-moving scaling dimension, so that:
\eqn\dimphi{\Delta(\Phi_{jm\bar{m}})={-j(j+1)\over k-2}~,}
as it should.
The operators $\Phi_{jm\bar{m}}$ are primaries of $\widehat{SL(2)}_L
\times \widehat{SL(2)}_R$.

Single valuedness of the wave function on $AdS_3$ leads to the constraint:
\eqn\mmim{m-\bar{m}\in {\bf Z}~.}
We will shortly discuss how to implement this condition.

Next we discuss to which $SL(2)$ representations the operators
$\Phi_{jm\bar{m}}$ belong.
Unitarity of the $SL(2)/U(1)$ theory implies that $j$ belongs
to one of the following regions \gk\ (for early work on unitarity
of $SL(2)/U(1)$ see \early):
\eqn\jinr{j\in {\bf R}~,\qquad -\frac12<j<{k-3\over 2}~, \qquad
\Delta(\Psi_{jm\bar{m}})\geq 0~,}
or
\eqn\jinir{j\in -\frac12+i{\bf R}~,}
where $\Delta(\Psi_{jm\bar{m}})$ is given in \dimpsi.
When the $SL(2)/U(1)\times S^1_X$ is recomposed into an $SL(2)$,
operators with $j$ in the ranges above fall
into three kinds of representations of $SL(2)$ \mo\
(see \early\ for earlier works).
Operators in the range \jinr\ fall into either
lowest weight principal discrete representations:
\eqn\mjp{\DD^+_j:\qquad m=j+n~, \qquad n=1,2,...~,}
or highest weight principal discrete representations:
\eqn\mjm{\DD^-_j:\qquad m=-(j+n)~, \qquad n=1,2,...~.}
Operators with $j$ in the range \jinir\ fall into
principal continuous representations of $SL(2)$:
\eqn\ma{\CC_{j,\mu}:\qquad
m=\mu~, \mu\pm n~,\qquad 0\leq \mu<1~,\qquad n=1,2,...~.}
Let us now comment on the spacetime interpretation of vertex operators
$V(j;m,\bar{m})$, which are primaries of $\widehat{SL(2)}$,
built from the various $SL(2)$ representations above. The principal
continuous representations $\CC_{j,\mu}$ are associated with tachyonic
states. On the other hand,
local observables correspond to vertex operators $V(j;m,\bar{m})$
in the highest weight representations $\DD_j^-$
and lowest weight ones $\DD_j^+$ of $SL(2)$. 
They correspond in spacetime to the modes
$A^h_{m\bar{m}}$ of a primary field with weight $h=j+1$.
To see that, recall that the commutator of a physical vertex operator
with the spacetime Virasoro generators is given by \gks:
\eqn\comm{[L_n,V(j;m,\bar{m})]=(nj-m)V(j;m+n,\bar{m})~,}
while in a two dimensional CFT:
\eqn\commcft{[L_n,A^h_{m\bar{m}}]=(n(h-1)-m)A^h_{m+n,\bar{m}}~.}
In particular, the operators in $\DD_j^-$ are the creation operators,
generating ingoing states from the spacetime vacuum,
while those in $\DD_j^+$ are their hermitian conjugates.

\subsec{The ``Twist Field" and the New Operators}
Above we remarked that single valuedness of the wave function on
$AdS_3$ space requires \mmim.
In \refs{\gks,\ks} the condition \mmim\ was imposed ``by hand.''
To obtain this condition ``intrinsically'' in the $AdS_3$
worldsheet theory, we need to include ``twist fields''
$t^w_{sl(2)}$ which impose such a condition via mutual locality.

Let us stress that we are not twisting an otherwise consistent theory,
rather it is consistency of the theory 
(tree level unitarity and higher loop modular invariance)
which requires the inclusion
of these twist fields, and of all the twisted operators which arise through
OPEs with the twist fields.

We thus introduce the operator:
\eqn\ttw{t^w_{sl(2)}(z,\bar z)=t^w_{sl(2)}\bar{t}^w_{sl(2)}=
e^{-w(\int  J^3(z)+\int  \bar{J}^3(\bar{z}))}=
e^{w\sqrt{k\over 2}(X(z)+\bar{X}(\bar{z}))}~, \qquad w\in {\bf Z}~.}
The OPE of $t^w_{sl(2)}$ with $\Phi_{jm\bar{m}}$ takes the form
\eqn\ope{\eqalign{t^w_{sl(2)}(z,\bar z)\Phi_{jm\bar{m}}(z',\bar z')&\sim
(z-z')^{-wm}(\bar z-\bar z')^{-w\bar m}\Phi_{jm\bar{m}}^w(z',\bar z')\cr
&=(z-z')^{-(m-\bar m)w}|z-z'|^{-2\bar m w}\Phi_{jm\bar{m}}^w~,}}
where
\eqn\phiw{\Phi_{jm\bar{m}}^w=\Psi_{jm\bar{m}}
e^{\sqrt{2\over k}[(m+{k\over 2}w)X+(\bar m+{k\over 2}w)\bar X]}~.}
{}From \ope\ we see that mutual locality with $t^w_{sl(2)}$ indeed
implies the condition \mmim. Now, consistency of the theory
implies that we must include also the ``twisted'' operators
$\Phi_{jm\bar{m}}^w$ \phiw.

The scaling dimension of $\Phi_{jm\bar{m}}^w$ is:
\eqn\dimphiw{\Delta(\Phi_{jm\bar{m}}^w)=
\Delta(\Psi_{jm\bar{m}})-{(m+\frac{k}2w)^2\over k}=
{-j(j+1)\over k-2}-{k\over 4}w^2-mw~.}
Considering now a vertex operator $V^w$ defined by:
\eqn\physop{V^w=V_\Delta \Phi_{jm\bar{m}}^w~,}
where $V_\Delta$ is a primary 
operator in the CFT on $\NN$ with weight $\Delta$,
the physicality condition is:
\eqn\physcon{-{j(j+1)\over k-2}-{k\over 4}w^2-mw+\Delta=1~.}
To find the spacetime properties of $V^w$ we first need to recall
the following OPE (see \gks\ for details):
\eqn\jgn{J^3(z)\gamma^n(z')\sim {n\gamma^n(z')\over z-z'}~.}
This and \jthree\ imply that\foot{$\gamma^m$ is the free field representation
of $\Phi_{0m}$ (see \gks).}:
\eqn\gx{\gamma = \Psi_{\gamma}e^{\sqrt{\frac2{k}}X}~,}
and therefore, using \phiw\ and the fact that
$\gamma^n$ has a regular OPE with $\Phi_{jm}$ \phipsi,
we find that
\eqn\gphiw{\gamma^n(z)\Phi^w_{jm}(z')\sim (z-z')^{-nw}O(z')~,}
where $O$ is some operator.
Second, using \jthree, \jpm, \phiw, and the standard OPEs in the
parafermionic theory (in its free field representation), we find:
\eqn\opes{\eqalign{
J^3(z)\Phi_{jm}^w(z')&\sim {(m+\frac{k}2 w)\Phi_{jm}^w(z')\over z-z'}~,\cr
J^{\pm}(z)\Phi_{jm}^w(z')
&\sim (z-z')^{\mp w-1}(m\mp j)\Phi_{j,m\pm 1}^w(z')+\dots~.}}
Hence, from \ln, \gphiw, \opes, we find that for $w<0$:
\eqn\lnv{[L_0,V^w]=-(m+\frac{k}2w)V^w~, \qquad
[L_n,V^w]=0~, \qquad n\geq 1~.}
{}From \lnv, one learns that the spacetime state $|V^w\rangle$ 
created by $V^w$ from the vacuum obeys:
\eqn\obeys{L_0|V^w\rangle=h|V^w\rangle~, \qquad
L_n|V^w\rangle=0~, \qquad n\geq 1~,}
where
\eqn\hhh{h=-(m+\frac{k}2w)~.}
Hence $|V^w\rangle$ is a primary state with weight $h$.
Since $V^w$ obeys \lnv, rather than \comm, it corresponds in spacetime
to the mode $m_{st}=-h$ of a primary field plus something which
annihilates the spacetime vacuum.
The latter piece is presumably an indication of the non-locality 
of the $w\neq 0$ sectors\foot{In local CFTs operators cannot obey \lnv.
However, we expect that the spacetime CFT corresponding to string theory on
$AdS_3$ with a vanishing RR background is a non-local (singular) CFT \sw.}, 
due to its relation to long strings \mo.
The part of $V^w$ corresponding to a mode of a primary, on which we focus,
is however a local observable.

Solving \physcon\ for $m$ in \hhh, we get:
\eqn\flow{h={k|w|\over 4}+{1\over |w|}\left(-{j(j+1)\over k-2}+\Delta-1
\right)~,}
in agreement with \mo.

\noindent
A few comments are in order:
\item{1.}
Equation \lnv\ implies that all operators with $w<0$ correspond to
creation operators, up to something which annihilates
the vacuum. Namely, also operators $\Phi_{jm}$ in the
$\DD_j^+$ representation, which correspond to annihilation operators, when
twisted with $w<0$ create incoming states from the spacetime vacuum.
This is not a surprise
since in the spectral flow picture \mo, the $w=-1$ flow of the $\hat{\DD}_j^+$
representation  of $\widehat{SL(2)}$ is nothing else than
$\hat{\DD}_{\tilde{j}={k\over 2}-j-2}^-$.
\item{2.}
{}For $w=0$, the spacetime primary states are created from the vacuum by 
the operators corresponding to $\Phi_{j,-j-1}$,
which is the highest weight of $\DD_j^-$. On the other hand, \lnv\ implies
that $\Phi_{jm}^w$ for any $m$ (compatible with \mjp\ or \mjm) create
from the vacuum
a primary state. This again can be understood using the spectral flow
picture, where one finds that all the primary states of $\widehat{SL(2)}$,
which sit in a $\DD_j^\pm$ representation of $SL(2)$, flow to states in
the representation $\hat{\DD}^{\pm,w}_j$ (in the notation of \mo)
that are all highest weights of $SL(2)$ (note that $w<0$).
The other states in this $SL(2)$ representation correspond to twists
of $\widehat{SL(2)}$ descendants.
\item{3.}
The above discussion can be repeated for $w>0$, exchanging ingoing for
outgoing states. This is compatible with the picture where $w$ is
related to a (non-preserved) winding number 
of long strings near the boundary of $AdS_3$, 
since winding of opposite sign signals charge conjugation.

\newsec{The Superstring on $AdS_3$}

Let us first recall the new ingredients involved in a fermionic string
on $AdS_3$ relative to the bosonic case.
This theory has affine $SL(2)$ {\it super}currents
\eqn\scur{\psi^a+\theta \sqrt{\frac2k} J^a ~,\qquad a=1,2,3~,}
where
\eqn\jA{J^a=j^a-\frac{i}{2}\epsilon^a{}_{bc}\psi^b \psi^c~,}
and
\eqn\norms{\eqalign{\psi^a(z)\psi^b(z')
&\sim\frac{\eta^{ab}}{z-z'}~,
\qquad \eta^{ab}={\rm diag}(+,+,-)~, \cr &\cr
J^a(z)J^b(z')
&\sim\frac{\frac{k}2\eta^{ab}}{(z-z')^2}+
\frac{i\epsilon^{ab}{}_cJ^c}{z-z'}~.}}
The purely bosonic currents $j^a$ generate an affine $SL(2)$
algebra at level $k+2$ and commute with $\psi^a$, whereas
the total (physical) currents $J^a$ generate a level $k$ $SL(2)$
algebra\foot{From now on we denote by $k$ the total
level of $SL(2)$, while in the previous section it denoted
the bosonic level.} and act on $\psi$ as follows from \jA,\norms.
Therefore, the ``untwisted'' operators $\Phi_{jm\bar{m}}$
have scaling dimension:
\eqn\sdphi{\Delta(\Phi_{jm\bar{m}})={-j(j+1)\over k}~.}
The unitarity bound on $j$, written
in terms of the total level, is now \refs{\gk,\mo}:
\eqn\bound{-\half<j<{k-1 \over 2}~.}

\subsec{Review of the $N=2$ Superstring on $AdS_3\times \NN$}

To construct a {\it superstring} on $AdS_3\times \NN$ we need to
impose a chiral GSO projection. In the following we review the
construction of \refs{\gr,\bl} leading to $N=2$ supersymmetry in the 
boundary two dimensional CFT (the spacetime theory).
There are three conditions on the CFT background $\NN$:
\item{1.}
$\NN$ is a SCFT with central charge
\eqn\cn{c_{\NN}=15-c_{sl(2)}=15-\left({9\over 2}+{6\over k}\right)
={21\over 2}-{6\over k}~,}
so that the total worldsheet central charge on $AdS_3\times \NN$ is
critical: $c=15$.
\item{2.}
$\NN$ has an affine $U(1)$ which we write in terms of
a canonically normalized scalar $Y$ as:
\eqn\uy{J^Y=i\d Y~;}
we denote its worldsheet fermionic superpartner by $\chi^Y$.
\item{3.}
The quotient CFT $\NN/U(1)$, where the $U(1)$ we gauge is with
respect to $J^Y$ \uy, has an $N=2$ supersymmetry.
Let $J_R^{\NN/U(1)}$ be the $U(1)_R$ current of the $N=2$
worldsheet superconformal algebra of $\NN/U(1)$. We can write it in terms of
a canonically normalized scalar $Z$ as:
\eqn\ur{J_R^{\NN/U(1)}=i\sqrt{c_{\NN/U(1)}\over 3}\d Z=
i\sqrt{3-{2\over k}}\d Z~.}

\noindent
We introduce two additional canonically normalized scalars $H_1, H_2$,
the bosonizations of the $SL(2)\times U(1)_Y$ fermions:
\eqn\hh{\d H_1=\psi^1\psi^2~, \qquad i\d H_2=\psi^3\chi^Y~.}
The {\it spacetime} supercharges are constructed as \fms:
\eqn\gspace{G^\pm_r=(2k)^\frac14\oint dz\,
e^{-\frac\varphi2} S^\pm_r~,\qquad r=\pm\frac12~,}
where $\varphi$ is the scalar field arising in the bosonized
superghost system of the worldsheet supersymmetry, and the spin
fields $S^\pm_r$ are (see appendix A for the details of the computation):
\eqn\splus{\eqalign{ S^+_r & =
e^{-ir(H_1+H_2)-\frac{i}2\sqrt{3-\frac2k}Z+i\sqrt{\frac1{2k}}Y}~,\cr
 S^-_r & =
e^{-ir(H_1-H_2)+\frac{i}2\sqrt{3-\frac2k}Z-i\sqrt{\frac1{2k}}Y}~.}}
The algebra generated by the supercharges is
\eqn\stalg{\eqalign{\{G^+_r,G^-_s\} & =2L_{r+s}+(r-s)J_0
~,\qquad r,s=\pm\frac12\cr
[L_m,L_n] & = (m-n)L_{m+n}~,~~~\,\qquad m,n=0,\pm1 \cr
[L_m,G^\pm_r] & = (\frac{m}2-r)G^\pm_{m+r} \cr
[J_0,G^\pm_r] & =\pm G^\pm_r}}
with all other (anti-)commutators vanishing. Up to picture-changing \fms,
$L_0,L_{\pm1}$ are given by \sltwo\ (with $J^a$ being the total currents), 
while $J_0$ is given by:
\eqn\stj{J_0=\oint dz J^R(z) = \sqrt{2k}\oint dz J^Y(z)~.}
The algebra \stalg\ is a global spacetime $N=2$ superconformal algebra
whose $U(1)_R$ generator is $J_0$.\foot{Here we took $J_0\rightarrow -J_0$
and $S^+\leftrightarrow S^-$ relative to \gr, to be compatible with the
conventions of \gkp\ used later.}

Commuting $L_n$, the generators of the full spacetime Virasoro algebra \vir,
with the generators of the global algebra \stalg\ gives a full spacetime
$N=2$ superconformal algebra in the spacetime NS sector with
modes $G^\pm_r$,
$r\in {\bf Z}+\frac12$ and $J_n$, $n\in {\bf Z}$. Physical states are
constructed using physical vertex operators that are local with
respect to the supercharges \gspace; this is the analogue of the usual GSO
projection.

\subsec{The Supersymmetric ``Twist Field"}

To obtain a consistent superstring theory on $AdS_3$ we need to include a
twist operator which imposes the single valuedness condition \mmim.
The twist field $t_{sl(2)}^w$ in \ttw, with $J^3$ being the {\it total}
current,
is not good because it is not mutually local with the spacetime supercharges
$G_r^{\pm}$ \gspace--\splus.\foot{We could have used the bosonic twist
field $e^{-w\int j^3}$ which is mutually local with $G_r^{\pm}$, but
we find it natural to use the total $J^3$ since the spacetime Virasoro
generators \ln\ are constructed \gks\ from the total $SL(2)$ currents.}
A good twist operator $t^w$ which is mutually local with the
supercharges acts also on the $U(1)_Y$ factor of
$\NN\simeq \NN/U(1)\times U(1)_Y$. Explicitly, we include in the
superstring the operator\foot{{}From now on, for simplicity,
we only write the left moving part of the operators.}:
\eqn\tws{t^w=t^w_{sl(2)}t^w_{u(1)}=e^{-w\int J^3}
e^{\frac{w}2\int J^R}=e^{w\sqrt{\frac{k}2}(X+iY)}~,}
where $J^R$ is given in \stj, \uy.

Let us check that indeed $t^w$ is mutually local with the spacetime
supercharges. To do that, we use the facts:
\eqn\jjh{J^3=j^3-i\d H_1~,}
where $j^3$ is the bosonic current and $H_1$ is given in \hh,
hence,
\eqn\ejeh{e^{-\int J^3}\sim e^{iH_1}~,}
and
\eqn\ejno{e^{\frac12\int J^R}=e^{i\sqrt{\frac{k}2}Y}~.}
Using \tws, \ejeh, \ejno, and the form of the spin fields $S_r^{\pm}$
in \splus\ we find that
\eqn\tzsz{t^w(z)S_r^{\pm}(z')\sim (z-z')^{-(r\mp\frac12)w}O(z')~,}
where $O$ is some operator.
Since $r=\pm\frac12$ we get $r\mp\frac12\in {\bf Z}$ and, therefore,
$G_r^{\pm}$ survive the twist.

\noindent
Comments:
\item{1.}
The combinations $J^3\pm\frac12 J^R$ are null and thus, in particular,
\eqn\dimtw{\Delta(t^w)=0~.}
\item{2.}
As discussed before, we may decompose the $AdS_3\times \NN$ CFT
on $AdS_3\times S^1_Y \times \NN/U(1)$. Consistency with the GSO
projection \splus\ requires $S^1_Y$ to be a circle of radius
$R_Y=\frac1{\sqrt{k}}$ \gk\ (we work in the normalization where
the self-dual radius, obtained at $k=1$ \gk, is 1).
This means, in particular, that in the NS-NS sector an exponential factor
$e^{i(qY+\bar q\bar Y)}$ in vertex operators
has $q=\frac1{\sqrt{2}}(\frac{P}{R_Y}+WR_Y)$
and $\bar q=\frac1{\sqrt{2}}(\frac{P}{R_Y}-WR_Y)$, where $P,W\in {\bf Z}$.
Invariance under the twist imposes the condition
$(m-\bar m)+\sqrt{\frac{k}2}(q-\bar q) \in {\bf Z}$.
But in the NS-NS sector we have
$\sqrt{\frac{k}2}(q-\bar q)=\sqrt{k}WR_Y=W\in {\bf Z}$,
and therefore in the NS-NS sector only
operators obeying $m-\bar m\in {\bf Z}$ survive the twist.
In the R-NS sector, \splus\ implies that $q\to q\pm \sqrt{\frac1{2k}}$,
therefore, $\sqrt{\frac{k}2}(q-\bar q)=W\pm\frac12\in {\bf Z}+\frac12$,
which further implies $m-\bar m\in {\bf Z}+\frac12$.
This is consistent with the expectation that worldsheet operators
in the R-NS sector correspond to spacetime fermions.

\noindent
We chose to introduce in the theory the holomorphic twist operators
\eqn\tm{t^w_+=e^{w\sqrt{\frac{k}2}(X+iY)}~,}
where $J^3=-\sqrt{k\over 2}\d X$.
However, one could also choose to twist with their
complex conjugates
\eqn\tmm{t^w_-=e^{w\sqrt{\frac{k}2}(X-iY)}~.}
The two are related by the automorphism $a_{\pm}=e^{\pm i\sqrt{2k}Y}$
(which is mutually local with the supercharges \gspace, \splus,
and it changes the quantized momenta $P$ on $S^1_Y$ by an even number).
We will see that there is no ambiguity in considering both of the
above types of twist fields.

\newsec{Chiral Spectrum}
A property of two dimensional
theories with at least $N=2$ supersymmetry
is the existence of chiral states, which belong to short supermultiplets.
The spectrum of these operators does not change under continuous deformations,
since their weight is given in terms of their $R$-charge.
We now turn to study the chiral spectrum of the $N=2$
spacetime theory, as given by the superstring theory construction
of the preceding section. The details of the computations needed to obtain
the results of this section are contained in appendix A.

\subsec{The Untwisted Sector}
Starting from the NS sector of the worldsheet theory, there are two kinds of
vertex operators leading to spacetime chiral states.
We first consider the following vertex operators:
\eqn\vtac{\XX_R=e^{-\varphi}e^{iqY}V \Phi_{jm}~,}
where $V$ is an operator in the $N=2$ SCFT
on $\NN/U(1)$ with weight $\Delta_V$ and $R$-charge equal to $r_V$.
To simplify the discussion, we restrict to $r_V\geq 0$. The inclusion
of the $r_V\leq 0$ states amounts to complex conjugation of the
spacetime operators.

The operators \vtac\ have to satisfy the on-shell condition:
\eqn\phystac{-{j(j+1)\over k}+\half q^2 +\Delta_V=\half~,}
and the GSO projection:
\eqn\gsotac{q\sqrt{2\over k}-r_V \in 2{\bf Z} +1~.}
The operator \vtac\ has spacetime weight $h=j+1$ and spacetime
$R$-charge $R=\sqrt{2k}q$,
thus requiring chirality (or anti-chirality) in spacetime implies:
\eqn\chitac{j+1=h=\pm {R\over 2}=\pm \sqrt{k\over 2}q~.}
Putting together \chitac, \phystac\ and \gsotac\ (see appendix A for the
details),
we get the spacetime anti-chiral operators corresponding to the following
vertex operators \gkp:
\eqn\chix{\XX_{-k(1-r_V)}=e^{-\varphi}e^{-i\sqrt{2\over k}(j+1)Y}V
\Phi_{j={k\over 2}(1-r_V)-1,m}~,\qquad\qquad r_V+{1\over k}-1<0~,}
where the inequality is implied by \bound, and $V$ has to be a chiral
operator on the worldsheet, $\Delta_V={r_V\over 2}$.

The second set of operators is:
\eqn\vgrav{\WW_R=e^{-\varphi}e^{iqY}V (\psi\Phi_{j})_{j-1,m}~,}
where $(\psi\Phi_{j})_{j-1}$ is the combination of $\Phi_{j}$ with
the $SL(2)$ fermions that has $SL(2)$ spin $j-1$ 
(see \kll\ and appendix A for its explicit form).
These operators have to satisfy the on-shell condition:
\eqn\physgrav{-{j(j+1)\over k}+\half q^2 +\Delta_V=0~,}
and the GSO projection:
\eqn\gsograv{q\sqrt{2\over k}-r_V \in 2{\bf Z}~.}
Now the spacetime weight is $h=j$ and thus the condition for
spacetime (anti-)chirality is $j=\pm \sqrt{k\over 2}q$. Together with
\physgrav\ and \gsograv, we get the following spacetime chiral
operator:
\eqn\chiw{\WW_{kr_V}=e^{-\varphi}e^{i\sqrt{2\over k}jY}V
(\psi\Phi_{j={k\over 2}r_V})_{j-1,m}~,\qquad\qquad r_V+{1\over k}-1<0~.}
Again, $V$ is a chiral operator on the worldsheet ($r_V\geq 0$).

Moving to the Ramond sector of the theory, we find one additional
class of physical vertex operators which lead to chiral states in spacetime.
A general Ramond sector vertex operator will involve dressing operators with
spin fields similar to \splus. Note that the latter will not only
change the $SL(2)$ representation of $\Phi_j$ to $j\pm\half$, but
also alter the total charge with respect to both $J^Y$ and $J_R^{\NN/U(1)}$,
through their dependence on $Y$ and $Z$, respectively.
We thus consider the following operators:
\eqn\vram{\YY_R=e^{-{\varphi\over 2}}(S e^{iqY}V \Phi_{j})_{j-\half,m}^{q
+{1\over \sqrt{2k}}}~.}
The combination above is expanded in appendix A, where the derivation
of the following is also given.
The spin field in \vram\ is given by\foot{Note that $e^{-{\varphi\over 2}}
S_{\pm\half}$ on their own
are not mutually local with the supercharges \splus.}:
\eqn\spif{S_{\pm\half}=e^{\mp{i\over2}(H_1-H_2)-{i\over2}\sqrt{3-{2\over k}}Z
+i{1\over \sqrt{2k}}Y}~,}
the subscript being their $J^3$ eigenvalue.

The operators \vram\ satisfy the on-shell condition:
\eqn\physram{-{j(j+1)\over k}+\half q^2 +{q\over \sqrt{2k}}+\Delta_V
-{r_V\over 2}=0~,}
and the GSO projection:
\eqn\gsoram{q\sqrt{2\over k}-r_V \in 2{\bf Z}+1~.}
To derive \physram\ it is useful to write $V=ve^{i{r_V\over a}Z}$,
where $a=\sqrt{3-{2\over k}}$ and $v$ has a regular OPE with $J_R^{\NN/U(1)}$.
In addition, they are required to be BRST invariant.

The spacetime weight of \vram\ is $h=j+\half$ and their $R$-charge is given
by $R=\sqrt{2k}q+1$. Requiring (anti-)chirality in spacetime $j+\half=\pm(
\sqrt{k\over 2}q+\half)$, we end up with the following BRST invariant
vertex operator (which leads to a chiral state in spacetime):
\eqn\chiy{\YY_{1+k(r_V-1)}=e^{-{\varphi\over 2}}(S e^{i\sqrt{2\over k}jY}
V \Phi_{j={k\over 2}(r_V-1)})_{j-\half,m}~.}
As before, $V$ is chiral on the worldsheet with $r_V\geq 0$.
The bound \bound\ implies in this case:
\eqn\rbound{1<r_V+{1\over k}<2~.}

\noindent
Comments:
\item{1.}
In appendix A it is shown that one can also find BRST invariant vertex
operators in the Ramond sector which are anti-chiral in spacetime:
\eqn\chiyt{\tilde{\YY}_{1+k(r_V-2)}=e^{-{\varphi\over 2}}
(S^+ e^{-i\sqrt{2\over k}(j+1)Y} V \Phi_{j=-1+{k\over
2}(2-r_V)})_{j-\half,m}~,}
where $S^+$ is given in \splus\ and
the non-negative $r_V$ satisfies the same bounds \rbound.
However one can rearrange the $Y$ and $Z$ exponents to get:
\eqn\rearr{\tilde{\YY}_{1+k(r_V-2)}=e^{-{\varphi\over 2}}(S'
e^{-i\sqrt{2\over k}jY} \tilde{V}' \Phi_{j=-1+{k\over
2}(2-r_V)})_{j-\half,m}~,}
where:
\eqn\spifnew{S_{\pm\half}'=
e^{\mp{i\over2}(H_1+H_2)+{i\over2}\sqrt{3-{2\over k}}Z -i{1\over
\sqrt{2k}}Y}~,}
and $\tilde{V}'$ is a worldsheet anti-chiral operator which can be
written as:
\eqn\antichi{\tilde{V}'=ve^{{i\over a}(r_V-a^2)Z}~, \qquad\qquad
\Delta_V'=\Delta_V-r_V+{a^2\over 2}~,}
where again $a^2={c_{\NN/U(1)}\over 3}$.
We thus see that $\tilde{\YY}_{1+k(r_V-2)}$ is nothing else than the complex
conjugate of the operator $\YY_{1+k(r_V'-1)}$, where $r_V'=3-{2\over k}-r_V$
is the worldsheet $R$-charge of the complex conjugate of \antichi.
\item{2.}
The above formulas for the weight and charge of $\tilde{V}'$ are
exactly the ones given by spectral flow \ss\ in the $\NN/U(1)$ SCFT:
\eqn\etasf{\Delta_{\eta} = \Delta-r\eta + {c\over 6}\eta^2~, 
\qquad r_{\eta} = r - {c\over 3}\eta~,}
for $\eta=1$. This transformation
maps the NS sector of the theory to itself, and similarly for the Ramond
sector.
\item{3.}
The operators \chiy\ correspond to the spacetime bottom components
of the operators found in \gkp, which were shown to be top components
of a chiral superfield. One can indeed act with the relevant
spacetime supercharge on $\YY_{1+k(r_V-1)}$ and  find an operator
of the form \vtac, with the quantum numbers specified in 
subsection 4.1 of \gkp.

\noindent
To summarize, in the untwisted sector of the superstring on $AdS_3\times
\NN$, we find three families of vertex operators corresponding to
the bottom components of (anti-)chiral superfields in spacetime. For
convenience, we rewrite them below:
\eqn\allchi{\eqalign{
\XX_{-k(1-r_V)}&=e^{-\varphi}e^{-i\sqrt{2\over k}(j+1)Y}V
\Phi_{j={k\over 2}(1-r_V)-1,m}~,\qquad\qquad r_V+{1\over k}<1~,\cr
\WW_{kr_V}&=e^{-\varphi}e^{i\sqrt{2\over k}jY}V
(\psi\Phi_{j={k\over 2}r_V})_{j-1,m}~,\qquad\qquad r_V+{1\over k}<1~, \cr
\YY_{1+k(r_V-1)}&=e^{-{\varphi\over 2}}(S e^{i\sqrt{2\over k}jY}
V \Phi_{j={k\over 2}(r_V-1)})_{j-\half,m}~, \qquad\qquad 1<r_V+{1\over
k}<2~,}}
where $V$ is a chiral operator of $\NN/U(1)$ with an $R$-charge $r_V\geq 0$.
Their spacetime weights are in the following ranges:
\eqn\ranges{\eqalign{ \half &< h(\XX)={k\over 2}(1-r_V)\leq {k\over 2} \cr
0 & \leq h(\WW)={k\over 2} r_V < {k-1\over 2} \cr
0 &< h(\YY)=\half+{k\over 2}(r_V-1)<{k\over 2}~.}}

\subsec{The Twisted Sectors}
We now turn to consider the chiral spectrum in the spacetime CFT which
corresponds on the worldsheet to vertex operators obtained by twisting
the operators \vtac, \vgrav\ and \vram\ by the twist fields
$t^w_\pm$.
The result will be that each one of the (anti-)chiral operators
listed in \allchi\ gives rise upon twisting to a family of (anti-)chiral
operators whose spacetime weight is given by:
\eqn\patt{ h_w=h_0+{k|w|\over 2}~,}
where $h_0$ is the weight of the untwisted operators.
The appearance of this pattern will prove important later for
the identification of the spacetime theory.

In order to properly carry on the twist, it is useful to realize that
there are essentially three pieces in the twist fields \tws.
The $SL(2)$ part decomposes into a bosonic and a fermionic part,
according to:
\eqn\decom{J^3=-\sqrt{k\over 2}\d X=j^3-i\d H_1=-\sqrt{k+2\over 2}\d x
-i\d H_1~,}
where all the scalars are canonically normalized.
Therefore we can write the twist fields as:
\eqn\twt{t^w_\pm=e^{w\left(\sqrt{k+2\over 2}x+iH_1 \pm \sqrt{k\over 2}Y
\right)}~.}
Since the $\Phi_{jm\bar{m}}$ operators are purely bosonic, the expressions
\phipsi, \dimpsi\ and \dimphi\ apply,
but with $k$ replaced by $k+2$, the bosonic
level of $SL(2)$ in the supersymmetric context.
In the twisted operators however we take into account also the
piece containing $H_1$, so that their total weight is:
\eqn\sdimphiw{\Delta(\Phi_{jm\bar{m}}^w)=
{-j(j+1)\over k}-{k\over 4}w^2-mw~.}
On the other hand, the $U(1)_Y$ part of the twist will generate
the twisted operator $e^{i\left(q\pm w\sqrt{k\over 2}\right)Y}$ with
weight:
\eqn\twuone{ \Delta_{U(1)}=\half q^2 \pm qw\sqrt{k\over 2} +{k\over 4}w^2~.}
We are now ready to twist a general operator. Starting from \vtac\ and
acting with $t^w_\pm$, we get:
\eqn\twtac{\XX^w_R=e^{-\varphi}e^{i\left(q \pm w\sqrt{k\over 2}\right)Y}V
\Phi_{jm\bar{m}}^w~.}
The on-shell condition becomes:
\eqn\phystt{-{j(j+1)\over k}-mw+\half q^2 
\pm qw\sqrt{k\over 2}+\Delta_V=\half~,}
while the GSO condition stays \gsotac.
The spacetime weight is:
\eqn\hsusy{h=-m-{kw\over 2}~,}
where again $w<0$, and the $R$-charge is:
\eqn\rsusy{R=\sqrt{2k}q\pm kw~. }
Actually, had we straightforwardly twisted the physical operators
\chix\ by $t^w_+$, the physicality condition \phystt\
would boil down to imposing $m=-j-1$, giving rise to anti-chiral
states in spacetime labelled by $w$.

The fact that physical twisted fields corresponding
to (anti-)chiral operators in spacetime are the twist of (anti-)chiral
physical fields is not surprising, since the twist operators $t^w_\pm$
have worldsheet weight zero and are themselves ``(anti-)chiral," in the
sense that their $J^3$ eigenvalue is the absolute value of half of
their $J^Y$ eigenvalue.

{}From the above discussion we thus see that starting with an operator
of the list \allchi, taken to be in the $\DD^-_j$ representation
so that it creates an (anti-)chiral state in spacetime,
the twist of a chiral state with $t_-^w$ leads to a chiral state,
and the twist of an anti-chiral state with $t_+^w$ leads
to an anti-chiral state (remember that
we take $w<0$).

The vertex operators obtained by twisting \chix\ by $t^w_+$ are:
\eqn\chixw{\XX^w_{-k(1-r_V)}=e^{-\varphi}e^{-i\sqrt{2\over k}(j+1-{kw\over 2}
)Y}V \Phi^w_{j={k\over 2}(1-r_V)-1,m=-j-1}~,
\qquad\qquad r_V+{1\over k}-1<0~.}
They correspond to anti-chiral operators in spacetime with weight:
\eqn\weitx{h_w={k\over 2}(1-r_V)+{k|w|\over 2}=-{R\over 2}~.}
They clearly follow the pattern \patt\ with respect to the states
created by \chix.

The vertex operators obtained by twisting \chiw\ by $t_-^w$ are:
\eqn\chiww{\WW^w_{kr_V}=e^{-\varphi}e^{i\sqrt{2\over k}(j-{kw\over 2})Y}V
(\psi\Phi^w_{j={k\over 2}r_V})_{j-1,m=-j}~,\qquad\qquad r_V+{1\over k}-1<0~.}
Note that $(\psi \Phi_j)_{j-1,m=-j}=\psi^+ \Phi_{j,m=-j-1}$ (ignoring
pieces that annihilate the vacuum).
They correspond to chiral operators in spacetime with weight:
\eqn\weitw{h_w={k\over 2}r_V+{k|w|\over 2}={R\over 2}~.}
They also follow the pattern \patt, with respect to the states created
by \chiw.

Twisting by $t_-^w$ the vertex operators \chiy, one obtains:
\eqn\chiyw{\YY^w_{1+k(r_V-1)}=e^{-{\varphi\over 2}}(S
e^{i\sqrt{2\over k}(j-{kw\over 2})Y}
V \Phi^w_{j={k\over 2}(r_V-1)})_{j-\half,m=-j-\half}~, \qquad\qquad 1<r_V
+{1\over k}<2~.}
They correspond to chiral operators in spacetime with weight:
\eqn\weity{h_w=\half +{k\over 2}(r_V-1)+{k|w|\over 2}={R\over 2}~.}
They follow the pattern \patt, with respect to the operators \chiy.
More details on the derivation of \chixw, \chiww\ and \chiyw, and on
the discussion which follows, are to be found in appendix A.

The vertex operators \chixw, \chiww\ and \chiyw\ (together with their
charge conjugates) give rise to the chiral spectrum of the spacetime
theory obtained by the perturbative string analysis. Given the
chiral spectrum of the worldsheet CFT on $\NN/U(1)$, each
state being characterized by $r_V$, we get the vertex operators corresponding
to (anti-)chiral states in spacetime listed in \allchi. Each one of the
above operators is the $w=0$ member of a family of (anti-)chiral states
forming a regular pattern of weights increasing by ${k\over 2}$ each
time $w$ is increased by 1. The vertex operators corresponding to these
states are obtained by twisting the operators in \allchi\ by the appropriate
twist field.

In principle, one should also consider the possibility of
twisting an (anti-)chiral
vertex operator with the ``wrong" twist field, since this is a possibility
which is by no means excluded by the worldsheet CFT.
To verify that these are not additional chiral operators
we now show that vertex operators obtained in this way
just lead to a reshuffling between, and among, the different families above.

{}For instance, twisting
\chix\ by $t^w_-$, one can see that the physicality condition
\phystt\ will now impose $m=j+1$. We thus conclude that $t^w_-$ has
to act on $\XX$ operators in the $\DD_j^+$ representation.
The outcome is:
\eqn\chixo{\tilde{\XX}^w_{-k(1-r_V)}=e^{-\varphi}e^{-i\sqrt{2\over k}
(j+1+{kw\over 2})Y}V \Phi^w_{j={k\over 2}(1-r_V)-1,m=j+1}~,
\qquad\qquad r_V+{1\over k}-1<0~.}
The spacetime weight is now:
\eqn\weitil{h_w=-j-1-{kw\over 2}={k\over 2}(r_V-1) +{k|w|\over 2}={R\over
2}~.}
Note that the state is now chiral. The expression \weitil\ is of course
valid only for $w\leq -1$. Noting that $h_{-1}={k\over 2}r_V$, we see
that \weitil\ actually follows the pattern \weitw\ (with $|w|$ replaced
by $|w|-1$), that is, with respect to the operators \chiw.

This was already noted in the first remark after \flow, that the
$w=-1$ twist of an operator in the $\DD_j^+$ representation is an operator in
the $\DD_{\tilde{j}={k\over 2}-j-1}^-$ representation\foot{Note that we
have to replace $k$ in the first comment after \flow\ with $k+2$.}.
Namely in the case above
we have that $\hat{\DD}^{+,w=-1}_{j={k\over 2}(1-r_V)-1}=
\hat{\DD}^-_{\tilde{j}={k\over 2}r_V}$.
We can thus straightforwardly write:
\eqn\same{\eqalign{\tilde{\XX}^{w=-1}_{-k(1-r_V)}&
=e^{-\varphi}e^{i\sqrt{2\over k}
\tilde{j}Y}V e^{-iH_1}\Phi_{\tilde{j}={k\over 2}r_V, m=-\tilde{j}-1} \cr &
=e^{-\varphi}e^{i\sqrt{2\over k}\tilde{j}Y}V(\psi\Phi_{\tilde{j}={k\over 2}
r_V})_{\tilde{j}-1, m=-\tilde{j}}=\WW_{kr_V}~,}}
where we used the fact that $e^{-iH_1}=\psi^+$.

Similarly, acting with $t_+^{w=-1}$ on $\WW_{kr_V}$ one gets
$\XX_{-k(1-r_V)}$, and acting with $t_+^{w=-1}$ on $\YY_{1+k(r_V-1)}$
one gets $\tilde{\YY}_{1+k(r_V-2)}=(\YY_{1+k(r_V'-1)})^*$ (recall the
discussion after \antichi).
We thus conclude that there are no new twisted vertex operators leading
to (anti-)chiral states in spacetime, obtained by acting with the ``wrong"
twist field.

To summarize, twisting with $t_\pm^{w<0}$ the vertex operators in \allchi\
we obtain the following results:
\eqn\alltwi{\eqalign{t^w_+(\XX_{-k(1-r_V)})&=\XX_{-k(1-r_V)}^w~, \qquad
\qquad t^w_-(\XX_{-k(1-r_V)})=\WW_{kr_V}^{w+1}~, \cr
t^w_-(\WW_{kr_V})&=\WW_{kr_V}^w~, \qquad \qquad \qquad
t^w_+(\WW_{kr_V})=\XX_{-k(1-r_V)}^{w+1}~, \cr
t^w_-(\YY_{1+k(r_V-1)})&=\YY_{1+k(r_V-1)}^w~, \qquad \qquad
t^w_+(\YY_{1+k(r_V-1)})=(\YY_{-1-k(r_V-2)}^{w+1})^*~.}}
Two related remarks are the following:
\item{1.}
Assuming that the smallest spacetime CFT allowed by the formula \cst\
has $c=6k$ ($\MM_{6k}$ in \mkp),
we can now act on any chiral state in it with the $N=2$, 
$\eta=1$ spectral flow \etasf\ 
and obtain $h\rightarrow {c\over 6}-h=k-h$. The resulting operators
clearly cannot come from the untwisted $w=0$ sector -- the short string --
because of the bound \bound\ which forces $h\leq {k\over 2}$, see \ranges.
One therefore needs to include the long string sector to recover the spectrum
of the $c=6k$ theory.
Indeed we expect $\MM_{6k}$ to be the spacetime CFT of a theory of a single
long string, and the ``missing'' $N=2$ chiral states (roughly half of them)
could in principle ``disappear'' in the continuum (see the next remark).
\item{2.}
In the discussion of the untwisted sector
we did not consider the continuous representations $\CC_{j,\mu}$
of $SL(2)$, since they represent tachyons, and
are eliminated by the GSO projection. However in the $w\neq 0$ sectors such
representations are included in the physical spectrum; they correspond
to the continuum associated with the long strings, as shown in \mo.
{}For instance, the lowest lying physical operator surviving the GSO projection
is:
\eqn\cont{e^{-\varphi}\Phi^{w=-1}_{j,m}~,\qquad \qquad
j=-\half+is~, \qquad m={k\over 2}-{(k-1)^2\over 4k}
-{s^2\over k}~.}
(Recall that in this section $\Phi^{w}$ includes $e^{iw H_1}$, see near
eq.~\sdimphiw.)
In particular, the gap above which the continuum starts is $h=
{(k-1)^2\over 4k}$ (as is obtained from \cont\ with $s=0$
using \hsusy), in agreement with \sw. It is not clear if and
which of the operators in the continuum correspond to observables in
the spacetime theory. It is possible in principle
that some chiral operators could ``disappear'' in the continuum; this was
argued for the particular example of $AdS_3\times S^3\times T^4$ in \sw\
(we return to this example later), and
may also provide the missing chiral states discussed in the preceding remark.
Here our study is limited to local observables in the
discrete representations and their twists.

\newsec{The Candidate Spacetime Theory}

The analysis in the previous sections, using
perturbative worldsheet techniques, revealed that the chiral spectrum of the
spacetime CFT obeys a simple and regular pattern of duplication \patt.
We will show that the same pattern appears in a class of orbifold CFTs
-- symmetric products -- and we will thus argue that
the $c=6kp$ spacetime CFT dual to superstring theory
on a background of the general form $(AdS_3)_k \times\NN$ is (a
deformation of) a symmetric product CFT of the form:
\eqn\gen{(\MM_{6k})^p/S_p~.}
In what follows we will give a brief review of the subject of symmetric
product CFTs and discuss their chiral spectrum showing how the
pattern \patt\ naturally appears in this setup.

In order to compare the chiral spectrum obtained
from the worldsheet vertex operators -- corresponding to single string states
in spacetime -- to states in \gen,
we need to distinguish between single and multi-string
states from the spacetime perspective.
In the two dimensional CFT \gen, it is a priori
subtle to identify single particle states
due to the lack of asymptotic states and an S-matrix.
Yet, as shown below, we argue following \refs{\vw,\ms,\deboer}
that there is a natural identification of
``single particle states" with operators in certain twisted sectors.

According to the general theory of orbifold CFTs, modding by a
(discrete) non-Abelian symmetry group gives rise to twisted sectors
labelled by the conjugacy classes of the group.
In the case of a symmetric product $\MM^p/S_p$ \dmvv,
the conjugacy classes of $S_p$ are composed of disjoint cyclic permutations
of various lengths. Specifically, each conjugacy class can be
written as:
\eqn\perm{[C]=(N_1)^{n_1} (N_2)^{n_2} \dots (N_r)^{n_r}~,}
with $\sum_{i=1}^r n_i N_i=p$ and where by $(N)$ we denote
a single disjoint $Z_N$ cycle. Obviously conjugacy classes
come in one-to-one correspondence with partitions of the integer $p$.

It is natural to associate ``single particle states"
with operators in a single $Z_N$ twisted sector (more precisely,
corresponding to conjugacy classes of the form $[C_N]=(N)(1)^{p-N}$).
In particular, we also have the original single particle states of the
`diagonal' CFT $\MM_{diag}\cong \MM$.
Indeed, it turns out that the full (chiral) spectrum
of a symmetric product CFT can be represented in a Fock space \vw\ (when
$\MM$ is a $\sigma$-model on a complex manifold).
This Fock space is built of
elementary oscillators associated with (chiral) states of the kind
described above, i.e. chiral $Z_N$ twist fields and chiral operators in the
diagonal
CFT, which are therefore identified as single particle states.

It thus turns out that the generators of the chiral spectrum of
a symmetric product $\MM^p/S_p$ (corresponding to ``single particle
states") are the chiral operators in the $Z_N$ twisted sectors
(i.e. the sectors of maximal length) of cyclic orbifolds $\MM^N/Z_N$
with $N=1,2 \dots p$ (symmetrized by $S_p$).\foot{The 
$Z_1$ chiral spectrum is that of $\MM$ itself.}

\subsec{The Chiral Spectrum of Symmetric Products}

We now turn to computing the chiral spectrum of $\MM^p/S_p$
theories corresponding to single particle states (a similar
manipulation appeared in \deboer).
{}First we recall facts about the $\MM^N/Z_N$ orbifold (see appendix B
for the details).
Given an $N=2$ SCFT $\MM$, for any operator in $\MM$ with weight $h$ and
$R$-charge $R$, there is an operator in the $Z_N$ twisted sector
of $\MM^N/Z_N$ with weight and $R$-charge given by \refs{\klsc,\fks}:
\eqn\hrtwist{h^N={h \over{N}}+{c\over24}{N^2-1\over N}~, \qquad R^N=R~,}
where $c$ is the central charge of $\MM$.

Another property of $N=2$ theories is the spectral flow \ss\ which
relates any operator in the NS sector with weight $h_{NS}$ and $R$-charge
$R_{NS}$ to an operator in the Ramond sector with 
(see \etasf\ with $\eta=\half$):
\eqn\specflo{\eqalign{h_{R} &= h_{NS}-\half R_{NS} +{c\over 24}~, \cr
R_R&= R_{NS} - {c\over 6}~. }}
{}From the first line above one sees that chiral states of the NS
sector correspond under spectral flow to Ramond ground states.

Starting now with a chiral state in $\MM$ with
$h_{NS}=\half R_{NS}\equiv \half R$,
it flows to the Ramond ground state specified by:
\eqn\runtw{h_R={c\over 24}~, \qquad R_R= R-{c\over 6}~.}
Using \hrtwist\ to compute the weight and the charge
of the corresponding state in the $Z_N$ twisted sector, we obtain:
\eqn\rtwi{h^N_R={Nc\over 24}~, \qquad R^N_R= R-{c\over 6}~.}
Note that in all generality
the twist field corresponding to a Ramond ground state of $\MM$
is also a Ramond ground state, but now of $\MM^N/Z_N$.
This assures us that flowing back to the NS sector
we will end up with a chiral state of the orbifold CFT. Indeed, using
\specflo\ backwards and with $Nc$ instead of $c$, we get:
\eqn\chitw{h^N_{NS}=\half R +{c\over 12}(N-1)~, \qquad
R^N_{NS}=R +{c\over 6}(N-1)~.}
We conclude that any chiral state in $\MM$ gives rise to a chiral state in
$\MM^N/Z_N$.\foot{Note that applying \hrtwist\ directly to the original
chiral state would generically lead to a twisted state that is not
chiral.}

One can use the reasoning above backwards to show that any chiral state in
$\MM^N/Z_N$ corresponds in this manner to a chiral state in $\MM$. This
establishes a one-to-one correspondence between
chiral states in the building block theory $\MM$ (including the identity) and
chiral states in each of the $Z_N$ twisted sectors of $\MM^p/S_p$.

Recall that the spectrum obtained in \chitw\ corresponds to operators
in $\MM^p/S_p$ associated to single particle states.
Comparing the spectrum \chitw\ with $c=6k$ to the worldsheet result
\patt, we see that the chiral spectra follow the same pattern.
The operators in the $w=0$ sector on the worldsheet define chiral
operators of $\MM_{6k}$ with $h\leq {k\over 2}={c_\MM \over 12}$.\foot{See
the remarks at the end of section 4
on why we only see operators with $h\leq {k\over 2}$ and not $h\leq k$ as
expected.}$^{,}$\foot{The chiral spectrum
of $\MM_{6k}$ that we obtain from the worldsheet
has a symmetry of reflection around $h={k\over 4}$. The ``generator" of
this symmetry is actually the ``wrong" twist with $w=-1$ (see the
right column of \alltwi). It would be interesting to understand what
kind of spacetime operator would generate that symmetry. Note that this
cannot be a spectral flow \etasf\ for any value of $\eta$ (the symmetry
generated by the $\eta=1$
spectral flow is around $h={k\over 2}$).}
Identifying $|w|=N-1$, the pattern in \chitw\ coincides precisely
with \patt. For any operator in $\MM$ with weight $h$
which we find from the worldsheet, there is an operator in the $|w|$
twisted sector with weight
$h+{k\over 2}(N-1)$ associated to the $Z_N$ twisted sector of the
symmetric product.

\newsec{$N=4,3,2$ Examples}

We now illustrate the general structure presented in the previous sections in
a few concrete examples, some of which were already studied in the literature.

\subsec{Small N=4}

We start with the case of type II string theory on
$AdS_3 \times S^3 \times T^4$.
This background is described on the worldsheet by a product of 
WZW models $SL(2)_k \times SU(2)_k
\times U(1)^4$ and was shown to possess small $N=(4,4)$
superconformal symmetry in spacetime \gks.
As before, for simplicity we mainly discuss the left-moving sector.

In this case the spacetime $N=2$ subalgebra is constructed \gr\
as reviewed in section 3.1 with respect to
the $N=2$ worldsheet CFT $\NN/U(1)={SU(2)_k\over U(1)}\times
T^4$. Below we will first consider the $N=2$ spacetime chiral primaries
corresponding to the untwisted vertex operators \allchi; they actually
turn out to be identical to the list of spacetime $N=4$ chiral primaries
as given in \kll.

To construct the vertex operators \allchi, we first have to collect
the chiral operators of ${SU(2)_k\over U(1)}\times T^4$.
They are generated by:
\item{(i)}
The chiral operators of the $N=2$ minimal model ${SU(2)_k\over U(1)}$
with $R$-charge $r_i={i\over k}$
($i=0,\dots, k-2$).
\item{(ii)}
The chiral operators of the SCFT on $T^4$: two complex fermions $\psi^\pm$
with $R$-charge $r_\pm=1$, and the current $J$ bilinear in the
fermions, with $r=2$.

\noindent
The spacetime chiral operators constructed
purely from the minimal model satisfy
${r_i+{1\over k}<1}$, and thus give rise to operators
$\XX_{-k(1-r_i)}=\XX_{-(k-i)}$ and $\WW_{kr_i}=\WW_i$ in eq.~\allchi.
Taking products of the minimal model operators with the $\psi^\pm$
chiral operators of $T^4$, we obtain operators which satisfy
$1<r_\pm+r_i +{1\over k}<2$. They thus give rise to the vertex operators
$\YY^\pm_{1+k(r_\pm+r_i-1)}=\YY^\pm_{i+1}$.
Note that the operators $J$ do not give rise to any vertex
operator in \allchi\
since they do not fit in any one of the inequalities of \allchi.

To summarize, we get in the untwisted sector two towers of chiral states
from the NS sector, one with spacetime weight $h=0, \half, \dots, {k-2\over 2}$
and the other with $h=1, \dots , {k-1\over 2}, {k\over 2}$, and two
towers of states from the Ramond sector, both with weights
$h=\half, 1,\dots, {k-1\over 2}$.
This is the same list of vertex operators given in \kll.
Restoring $SU(2)_k$ notations, these are given by
(we parametrize the range of $j$ by an integer $i=0,\dots,k-2$, in
agreement both with the $SL(2)$ and the $SU(2)$ unitarity bounds):
\eqn\strchi{\eqalign{ \WW_j& =e^{-\varphi}V_j(\psi \Phi_j)_{j-1}~,
\qquad\qquad h=j={i\over 2}~, \cr
\YY^\pm_j&=e^{-{\varphi\over 2}} (S^\pm V_j \Phi_j)_{j+\half, j-\half}~, 
\qquad h=j+\half={i+1\over 2}~, \cr
\XX_j& =e^{-\varphi}(\chi V_j)_{j+1}\Phi_j~, 
\qquad\qquad h=j+1={i+2\over 2}~,}}
where $V_j$ is an operator of spin $j$ in the $SU(2)$ WZW and
$(\chi V_j)_{j+1}$ is its combination with the $SU(2)$ fermions which
has spin $j+1$.

To include the $w$ twisted sectors in a way that preserves not only
the spacetime $N=2$ superconformal symmetry but also the $N=4$ one, we
should keep those twisted vertex operators that are mutually local with
all the generators of the $N=4$ superalgebra.
Since in this case all the twist fields \tws\ are mutually local with the $N=2$
subalgebra and with the generators of the $SU(2)$ $R$-symmetry,
we must include all of  the $w$ twisted sectors. One then
finds that the above towers
continue and repeat themselves. Since the repetition pattern
\patt\ has a step of ${k\over 2}$, we see that every tower has one
state every $k$ ones that is missing, in an otherwise linear series
(we shall return to this point below).

String theory on $AdS_3\times S^3 \times T^4$ can be obtained
as the near horizon limit of $k$ NS5-branes and $p$ fundamental
strings. By U-duality it is related to a system of a single NS5-brane
and $kp$ fundamental strings \refs{\dijk,\sw}, so that they belong
to the same moduli space. Hence it is expected
\sv\ that the spacetime theory should be a
deformation of the $\sigma$-model on $(T^4)^{kp}/S_{kp}$ -- the instanton
moduli space of $kp$ strings in a 5-brane\foot{Actually, it is claimed 
\mmoores\ that in this particular case there is an extra $T^4$ attached in a
certain way to $(T^4)^{kp}/S_{kp}$. This extra piece is not ``seen'' by the
present worldsheet considerations.}.

The chiral spectrum of such a $(T^4)^{kp}/S_{kp}$
orbifold is obtained as follows.
The SCFT on $T^4$ has $N=4$ superconformal symmetry and its
$N=2$ chiral operators listed above are highest weights of
representations of the $N=4$  $SU(2)$ $R$-symmetry.
The $\psi^\pm$ are in the spin $\half$ and the $J$ is in the spin $1$
representation, their weight coinciding with their $SU(2)$ spin.
Applying the spectral flow procedure of section 5
to a chiral state of weight $j_r$ we
get in the $Z_N$ twisted sector a chiral state with weight $h^N=
j_r+{N-1 \over 2}$. This gives rise to four towers of single particle
chiral states, one with $h={N-1 \over 2}$, two with $h={N\over 2}$ and one 
with $h={N+1 \over 2}$, where $N=1,2,\dots$ ($N=1$ is the diagonal $T^4$).
Note that the above discussion is purely holomorphic. When combining
left and right movers one should match only twist fields corresponding
to the same $N$. Thus for a given $N$ one has a total of 16 possibilities,
6 with $h=\bar{h}$, 8 with $h=\bar{h}\pm {1\over2}$ and 2 with 
$h=\bar{h}\pm 1$.
These results can also be obtained by direct orbifold CFT analysis \orbif,
or by using the Poincar\'e polynomial \refs{\vw,\ms,\deboer}.

Since our general expectation is that the spacetime theory should be
a (deformation of a) symmetric product $\MM^p/S_p$, it is tempting to 
argue that
$(T^4)^{kp}/S_{kp}$ lies in the same moduli space as $((T^4)^k/S_k)^p/S_p$.
Indeed they have the same chiral spectrum. Then we would conclude
that $\MM_{6k}$ is in the moduli space of $(T^4)^k/S_k$.
This is not accidental and actually meshes nicely with the
general picture. As explained above, string
theory on $AdS_3\times S^3 \times T^4$ is in the moduli space of the
near horizon theory of a single NS5-brane and $kp$ fundamental strings.
As such, the spacetime theory is expected to be not only in the moduli space
of $(\MM_{6k})^p/S_p$, but also in the one of $(\MM_6)^{kp}/S_{kp}$, leading
to the conclusion that $\MM_{6k}$ is in the moduli space of $(\MM_6)^k/S_k$.

However, we now encounter an apparent difficulty since the spectrum computed
from string theory in the untwisted sector \strchi, which we would associate
to the spectrum of $\MM_{6k}$, is actually not exactly the one of
$(T^4)^k/S_k$ since we miss the states corresponding to the $Z_k$ twisted
sector ($i=k-1$ in \strchi).

It was argued in \sw\ that these ``missing" spacetime
chiral states might be found within the continuum, namely, from the vertex
operators constructed with continuous representations of $SL(2)$ (i.e. from
the long strings). The missing chiral states disappear in the
continuum because the background we consider, with vanishing RR fields,
is a singular locus of the moduli space, where the spacetime CFT is
expected to be non-local\foot{Note that the missing $N=4$ chiral primary in
each tower is the first among a series of $N=2$ chirals of $(T^4)^k/S_k$
with ${k\over 2}\leq h\leq k$, which are expected to ``disappear'' in 
the continuum (see the remarks at the end of section 4).}.

A possibly related difficulty of the worldsheet approach is that we cannot
consider the case where we have a single NS5-brane, that is when
$k=1$, since $k\geq 2$ by unitarity of the $SU(2)$ WZW model.

\subsec{Large N=4}

We now turn to the case of type II string theory on
$AdS_3 \times S^3 \times S^3 \times S^1$ \efgt\ which was shown
to have the large $N=(4,4)$ superconformal algebra.
This background is described on the worldsheet by a product of 
WZW models $SL(2)_k \times SU(2)_{k'} \times SU(2)_{k''}\times U(1)$. 
We will restrict to the case $k'=k''=2k$.

The untwisted chiral spectrum \allchi\ in this case can be constructed
similarly to the previous example,
where now $\NN/U(1)={{SU(2)\times SU(2)} \over U(1)} \times U(1)$ \gr.
Unlike the small $N=4$ case, in the
present case only part of the $N=2$ chiral operators are chiral
primaries of the large $N=4$ superalgebra. The latter consist of \refs{\efgt,
\ofer}:
\eqn\unscp{\eqalign{ \WW_j &=e^{-\varphi}V'_j V''_j(\psi \Phi_j)_{j-1}~,
\qquad\qquad h=j={i\over 2}~,\cr
\YY_j &=e^{-{\varphi\over 2}} (S V'_j V''_j\Phi_j)_{j+\half,
j+\half,j-\half}~, \qquad h=j+{1\over 2} ={i+1\over 2}~.}}
Operators of the form $\XX_j$ in \strchi\ do not
belong to the large $N=4$ chiral ring.
Note that in the above the unitarity bound \bound\ on the $SL(2)$ spin $j$ 
implies a stricter bound $j<{k'-2\over 4}$ on the $SU(2)$ spin
than required by unitarity of the $SU(2)$ WZW model.

Twisting the operators \unscp\ with \tws, we should keep only the
resulting operators which are mutually local with the full large $N=4$
superalgebra. Unlike the small $N=4$ case, only the $w\in 2{\bf Z}$
twisted operators preserve the large supersymmetry\foot{Note that there
will still be operators in the $w$ odd sectors, but not chiral ones.}.
Therefore in the chiral spectrum we obtain the pattern:
\eqn\evenpatt{h_w=h_0 +kn=h_0+{k'\over 2} n~, \qquad \qquad n={|w|\over 2}
\in {\bf Z}~.}
The $AdS_3 \times S^3 \times S^3 \times S^1$ background is obtained as the near
horizon limit of two orthogonal stacks of $k'$ NS5-branes and $p$ fundamental
strings along their intersection. As in the previous subsection, it can
be shown that this background is in the moduli space of the background
obtained from $k'p$ fundamental strings at the orthogonal intersection of 
two single NS5-branes \malpri. Hence, similar to the small $N=4$ case
one expects that $\MM_{6k}$
is in the moduli space of $(\MM_3)^{k'}/S_{k'}$. In particular,
$(\MM_{6k})^p/S_p$ is in the moduli space of $(\MM_3)^{k'p}/S_{k'p}$.
Indeed, it was argued in \efgt\
that the chiral spectrum \unscp\ is compatible with the one of a symmetric
product of a SCFT $\MM_3$ which is a $c=3$ theory of one scalar $X$
and its four fermionic superpartners $\psi^a$, and which possesses large $N=4$
superconformal  symmetry.

The $R$-symmetry representation of operators in $(\MM_3)^{k'p}/S_{k'p}$
is denoted by $SU(2)\times SU(2)$ spins $(j',j'')$.
Studying the $N=4$ chiral ring we focus on $j'=j''=j_r$ \dBps.
The chirality condition in this case is $h={1\over2}(j'+j'')=j_r$.
In the $\MM_3$ model there is just one chiral operator
coming from the $SO(4)$ vector of the fermions $\psi^a$.
Starting with a chiral state in $\MM_3$ with $h=j_r$ and applying the
($N=2$) spectral flow transformation of section 5 we get in the
$Z_N$ twisted sector a chiral state with $h^N=j_r+{N-1 \over 4}$.
Clearly, only when $N$ is odd $h^N\in \half {\bf Z}$ and
the twisted operators are in the chiral ring (when $N$ is even, $h^N \in
\half {\bf Z} +{1\over 4}$ which corresponds to $j_N'=j_N''\pm \half$,
see \efgt).
Thus we find two towers of chiral single particle states,
one with $h={N-1 \over 4}={i\over 2}$ and the other with
$h={N +1 \over 4}={i+1\over 2}$, for $N=2i+1$ with $i=0,1,2,\dots$.

{}Finally, we should mention that as in the previous case the
last chiral operators of $\MM_{6k}$ (and hence of each $Z_N$
twisted sector of $(\MM_{6k})^p/S_p$)
are missing in the worldsheet picture, with respect to the
$(\MM_3)^{k'}/S_{k'}$ candidate. Presumably they ``disappear" in the
continuum as in the small $N=4$ case \sw.

\subsec{N=3}

In \ags\ superstring theory on a background of the form
$AdS_3 \times G/H$ was considered, and shown to possess $N=3$ superconformal
symmetry in the dual spacetime CFT in two cases:
$AdS_3 \times SU(3)/U(1)$ and $AdS_3 \times SO(5)/SO(3)$.
In \spec\ the $N=3$ chiral spectrum of the dual CFTs is analyzed
using perturbative worldsheet methods, in the untwisted sector.

The physical vertex operators found in \spec\ are:
\eqn\psivi{e^{-\phi}U_{[r,s],j'={r+s\over 2}}( \psi
\Phi_{j={r+s\over 4}})_{j-1,m}~, \qquad \qquad h=j={r+s\over 4}~,}
where the $G/H$ part of the vertex operator
is labeled by the representation $[r,s]$ of $G=SU(3),SO(5)$
and the representation $j'$ of the unbroken $SU(2)$, which
coincides with the spacetime $R$-symmetry.
We also have that $j<{k-1\over 2}$ so that $j'<{k'\over 4}-1$, where
$k'=4k$ is the level of $SU(3)$ or $SO(5)$ (and of the unbroken $SU(2)$).
The above operators
are chiral in spacetime since $h={j'\over 2}$. For a given $R$-spin $j'$,
we find $2j'+1$ distinct chiral operators in the NS sector.
There are no physical vertex operators coming from the Ramond
sector giving additional spacetime $N=3$ chiral operators.

Again the physical vertex operators \psivi\
fall into the classification of \allchi\
with $\NN/U(1)={SU(3) \over U(1)^2}$ or $\NN/U(1)
={SO(5) \over {U(1) \times SO(3)} }$
where \psivi\ is of the type $\WW$, and the vertex operators of the form
$\XX,\YY$ give rise to chirals of $N=2$ that are not chirals of $N=3$.
Including the twisted sectors, only the even $w$ twisted operators
are $N=3$ chiral.

Unlike the $N=4$ examples, $\MM_{6k={3\over 2}k'}$ cannot be itself
in the moduli space of some symmetric product
of a smaller building block, even though arithmetically
it could be of the form $(\MM_{3\over 2})^{k'}/S_{k'}$.
The reason is that the chiral spectrum has
for a specific $R$-charge $j'$ a multiplicity of $2j'+1$, while in
a symmetric product the multiplicity at every level is
bounded from above by the total number of chiral operators in the building
block theory, as it can be deduced from \chitw.

Note also that the arguments of \sw\ in favor of a spacetime CFT based
on a more general $S_{k'p}$ orbifold
rely on U-duality and the simple
brane picture of the $N=4$ cases and possibly do not apply here.

\subsec{N=2}
As a last example, we consider the family of ``non-critical"
superstrings on $AdS_3\times \NN$, where $\NN/U(1)={SU(2)_n\over U(1)}$
is an $N=2$ minimal model at level $n$ (see section 4.1 of \gkp).
The relation between $n$ and the level of the $SL(2)$ WZW model is:
\eqn\kmin{k={n\over n+1}~.}
The chiral operators of $\NN/U(1)$ have $R$-charge $r_i={i\over n}$,
with $i=0,1,\dots, n-2$. They satisfy $1<r_i+{1\over k}<2$, so that
they give rise to vertex operators of the kind $\YY_{1+k(r_i-1)}=
\YY_{i+1\over n+1}$. Note that the identity in spacetime is given
by an operator of the kind $\WW_0$ where the $j=0$ representation
is the $SL(2)$ identity. The vertex operators coming from the twisted
sectors are straightforwardly obtained.

In this case the spacetime theory is a deformation of $(\MM_{6n\over n+1})^p
/S_p$. The CFT $\MM_{6n\over n+1}$ cannot be in the moduli
space of a symmetric product, even arithmetically, because its central charge
is not linearly ``quantized."

\newsec{Discussion}

In this work we studied a large family of
$N=2$ superstrings on $(AdS_3)_k\times \NN$, with $g_s^2\sim 1/p$ \gs.
In particular, we argued that the spacetime theory
is in the moduli space of $(\MM_{6k})^p/S_p$ \mkp,
where $\MM_{6k}$ is the $c=6k$,
$N=2$ CFT whose Hilbert space corresponds to the worldsheet vertex operators
of short strings, and a single long string.
The short string vertex operators give rise to local observables in
$\MM_{6k}$. On the other hand, the long string contribution to the spectrum
of $\MM_{6k}$ is expected to consist of non-local observables -- the
continuum. This is compatible with the expectation \sw\ that string theory
on $AdS_3$ with a vanishing RR background gives rise to a (singular)
non-local CFT in spacetime.
 
Our study used standard perturbative worldsheet techniques.
Needless to say that those are reliable only in the weak $g_s$ limit,
namely, when $p\gg k$.
Strong string coupling effects become important for operators
whose spacetime weights $h$ approach the order of $p$.
Yet, although we cannot trust the perturbative worldsheet results
concerning such operators, the $AdS_3/CFT_2$ correspondence allows us to
speculate about some non-perturbative properties.
{}For instance, the fact that the spacetime central charge is $c_{st}=6kp$
implies that the chiral spectrum must be cut at $h=kp$.
This means that the spectrum corresponding to single
string states is cut at the $|w|=p-1$ twisted sector.
Adding $n$ string states with $n\leq p$ and with the maximal value
of the total $w$ equal to $p$ will give rise to the chiral spectrum of a
symmetric product with precisely $p$ copies of $\MM_{6k}$,
as allowed by $c_{st}=6kp$.

\lref\matrix{L.~Motl, ``Proposals on nonperturbative superstring 
interactions,'' hep-th/9701025; 
T.~Banks and N.~Seiberg, ``Strings from matrices,''
Nucl.\ Phys.\  {\bf B497} (1997) 41, hep-th/9702187;
R.~Dijkgraaf, E.~Verlinde and H.~Verlinde,
``Matrix string theory,'' Nucl.\ Phys.\  {\bf B500} (1997) 43,
hep-th/9703030.}

Moreover, ``extrapolating'' to small $p$,
the general picture described in this work still seems to make sense,
even though at present we cannot support it with direct computations.
{}For instance, 
for $p=1$ we claim that the spacetime theory \mkp\ is $\MM_{6k}$.
We expect that the $AdS_3\times \NN$ string background in the $p=1$ case
is constructed in the near horizon of a single fundamental string,
in which case it can emit at most a single long string.
Indeed, the spacetime theory $\MM_{6k}$ is defined to correspond
to a single long string and its short string excitations.
More precisely, the short string bulk excitations consist
of the $w=0$ sector, while the continuum in the $|w|=1$ sector (see
the remarks at the end of section 4) corresponds to the
long string \refs{\sw,\mo}. The $w=0$ discrete
states may be regarded as long string excitations emitted into the bulk \sw.

{}For $p>1$, the operators in the $Z_N$ twisted sectors of $(\MM_{6k})^p/S_p$
are associated with $N$ long strings, similar to ``matrix strings'' \matrix.
Indeed, from the worldsheet perspective, the $|w|=N-1$ twists of operators
in $\MM_{6k}$ are associated with $N$ long strings. 
The largest cyclic twisted sector $Z_p$ corresponds to the maximal number
of long strings $p$ -- the number of fundamental strings used to generate
the $AdS_3$ background.

In sections 6.1, 6.2, in the $N=4$ examples,
we saw that $\MM_{6k}$ itself is in the moduli
space of a symmetric product, $(\MM_6)^k/S_k$ and $(\MM_3)^{k'}/S_{k'}$,
respectively.
This is not accidental and meshes nicely with the general picture.
In these cases the moduli space includes a point corresponding to the
background constructed out of $P=kp$ and $k'p$ fundamental strings,
respectively, thus leading to a maximal number of $P$ long strings
and a spacetime CFT in the moduli space of a symmetric product of $P$
elementary blocks.
This property is related to the rich U-duality and the simple brane
picture associated with such high supersymmetry examples.
We also expect $\MM_{6k}$ to be in the moduli space of a symmetric product
in backgrounds obtained by orbifolding of the $N=4$ examples, for instance,
$AdS_3\times (S^3\times S^3\times S^1)/Z_2$ \yis.
It would be interesting to understand which generic backgrounds $\NN$
lead to $\MM_{6k}$ which are themselves
in the moduli space of a symmetric product.

{}Finally, we should mention that the pattern we find is not restricted
only to the chiral spectrum. Repeating the computation of section
5.1 for an operator in $\MM_{c=6k}$ with weight $h$ and $R$-charge $R$
such that:
\eqn\delt{h-\half R=\delta~,}
one finds in the $Z_N$ twisted sector of $\MM^p/S_p$ an operator
with:
\eqn\nonchi{h^N-\half R^N={\delta\over N}~, \qquad \qquad
R^N=R+{c\over 6}(N-1)~.}
The worldsheet theory leads to the same pattern: twisting an operator
in the $w=0$ sector which corresponds in spacetime to an operator
with $h$ and $R$ satisfying \delt, we get an operator with:
\eqn\ncws{h^w-\half R^w=f(\delta)~,\qquad \qquad
R^w=R+k|w|~,}
where $f(\delta)$ depends on the particular operator we twist, and
is such that $f(0)=0$. When $\delta>0$, generically $f(\delta)\neq 
{\delta\over N}$ because the 2-$d$ CFT dual to the superstring 
on $AdS_3 \times \NN$, although it is conjectured to be in the
moduli space of $\MM^p/S_p$, is not at the symmetric product point.

\bigskip
\noindent{\bf Acknowledgements:}
We are grateful to D.\ Kutasov for many valuable discussions and
useful comments. We also thank G.~Sarkissian for pointing out the 
references \refs{\klsc,\fks}.
This work is supported in part by the BSF -- American-Israel Bi-National
Science Foundation -- and by the Israel
Academy of Sciences and Humanities -- Centers of Excellence Program.
AG thanks the Theory Division at CERN for its hospitality.

\appendix{A}{Detailed Computation of Some Physical Vertex Operators}
We give here in detail some of the computations underlying the results
of sections 3 and 4. The main object is finding physical vertex 
operators, both in the NS and Ramond sectors, and both in the ``untwisted"
($w=0$) and ``twisted" sectors.

\subsec{The BRST Invariant Supercharges}
We briefly repeat here the derivation \gr\ of the physical $N=2$ spacetime
supercharges, using the notations of the present paper. 
In the process we introduce the worldsheet supercurrent
which we will also use later to impose BRST invariance of the Ramond
sector vertex operators.

The worldsheet $N=1$ supercurrent is given by:
\eqn\supcur{G_{tot}= G_{SL(2)} + G_{U(1)} + G_{\NN/U(1)}~,}
where, using the notation of section 3 and recalling that $\epsilon^{123}=1$:
\eqn\slcur{\eqalign{G_{SL(2)}&=\sqrt{2\over k}\left(\psi^a j_a - {i\over 6}
\epsilon_{abc}\psi^a \psi^b \psi^c\right)=\sqrt{2\over k}\left(\psi^a j_a
+i\psi^1 \psi^2 \psi^3 \right)~, \cr
G_{U(1)} &= \chi^Y J^Y = i \chi^Y \d Y~, }}
and we can decompose $G_{\NN/U(1)}$ according to the $N=2$ worldsheet
superconformal symmetry of $\NN/U(1)$, i.e.:
\eqn\noverucur{G_{\NN/U(1)}=G^+ + G^-=g^+ e^{i {Z\over a}} + 
g^- e^{-i {Z\over a}}~,}
where $Z$ is defined in \ur, 
$g^\pm$ have regular OPE with $Z$, and:
\eqn\adef{a^2=3-{2\over k}.}
To construct physical supercharges like \gspace\ we need to pick spin fields
$S$ such that $e^{-{\varphi\over 2}} S$ is BRST invariant, which translates
to requiring that there is no $z^{-3/2}$ singular term in the OPE
of $G_{tot}$ with $S$.

Using the bosonization \hh, a general spin field $S$ is given by:
\eqn\spigen{S_r^{\epsilon_2 \epsilon_Z}
\propto e^{-ir(H_1+\epsilon_2 H_2) + i \epsilon_Z ({a\over 2}Z-
{1\over \sqrt{2k}}Y)}~,}
where $r=\pm\half$, $\epsilon_2, \epsilon_Z=\pm 1$. The exact (relative)
prefactors and cocycle factors are determined as follows. 
The total $SL(2)$ currents after bosonization are:
\eqn\bososl{J^3=j^3-i\d H_1~, \qquad \qquad J^\pm=j^\pm\pm
c_1 c_2 e^{\mp iH_1}\left(e^{iH_2}-e^{-iH_2}\right)~,}
where $c_1, c_2$ are the cocycles associated to $H_1$ and $H_2$.
{}For the spin fields $S_r^{\epsilon_2 \epsilon_Z}$ to have the
OPEs with the $SL(2)$ currents appropriate to $j=\half$ and
$m=r=\pm \half$ (see for instance appendix A of \kll), 
we take the prefactors to be such that $S_r^{\epsilon_2 \epsilon_Z}$
take the form:
\eqn\pref{S_\half^{\epsilon_2 \epsilon_Z}=
e^{-{i\over 2}(H_1+\epsilon_2 H_2) + i \epsilon_Z ({a\over 2}Z-
{1\over \sqrt{2k}}Y)}, \qquad S_{-\half}^{\epsilon_2 \epsilon_Z}=
-\epsilon_2 c_1 c_2 e^{{i\over 2}(H_1+\epsilon_2 H_2) + i \epsilon_Z 
({a\over 2}Z- {1\over \sqrt{2k}}Y)}~.}
Since from \noverucur\ one sees that $G_{\NN/U(1)}$ has at most
a $z^{-1/2}$ singularity with the spin field \spigen, we focus
on the $SL(2)\times U(1)$ part of the worldsheet supercurrent, which
we can rewrite in a bosonized form:
\eqn\bosocur{\eqalign{G_{SL(2)} + G_{U(1)}=&{1\over \sqrt{k}}\left[
c_1 e^{iH_1}j^+ + c_1 e^{-iH_1} j^- - c_2\left(e^{iH_2}-e^{-iH_2}\right) j^3
\right] \cr
 & +{1\over \sqrt{k}}\left[ ic_2\left(e^{iH_2}-e^{-iH_2}\right) \d H_1 
+ic_2\left(e^{iH_2}+e^{-iH_2}\right) \sqrt{k\over 2} \d Y\right]~.}}
{}For finding BRST invariant supercharges, only the second line
of \bosocur\ gives potentially singular terms, and we can neglect all
the cocycles.

One then finds:
\eqn\brstspf{G_{tot}(z) S_r^{\epsilon_2 \epsilon_Z}(z') \sim
{1\over 2 \sqrt{k}}(z-z')^{-3/2}(-\epsilon_2-\epsilon_Z) S_r^{-\epsilon_2,
\epsilon_Z}(z')+\dots~.}
We thus conclude that the physical supercharges are built from spin fields
such that ${\epsilon_2 \epsilon_Z=-1}$, which is the result \splus.
Note that in this way the BRST condition also fixes the GSO projection
on the supercharges, which amounts to fixing the sign of 
$\epsilon_2 \epsilon_Z$.

\subsec{Untwisted Sector Physical Vertex Operators}
We now look for physical vertex operators leading to (anti-)chiral
operators in spacetime.
We start from the NS sector, and show that vertex
operators like \vtac: 
\eqn\nsvop{\XX=e^{-\varphi}e^{iqY}V\Phi_{jm}~,}
only lead to anti-chiral operators in spacetime.
Here we take $V$ to be an operator in $\NN/U(1)$ with $R$-charge $r_V\geq 0$,
so that it can be rewritten as $V=ve^{i{r_V\over a}Z}$.
In order for \nsvop\ to survive the GSO projection, it must be mutually
local with the supercharges \gspace--\splus. This imposes
$-\half \pm {r_V \over 2} \mp {q\over \sqrt{2k}} \in {\bf Z}$, which
is \gsotac.
Physicality is then obtained imposing the on-shell condition \phystac.
Since we are interested in (anti-)chiral operators in spacetime, we further
impose that the spacetime weight of \nsvop\ equals half of the absolute
value of its spacetime $R$-charge, leading to \chitac.
Moreover, quantities appearing in \phystac, \gsotac\ and \chitac\ must
satisfy two inequalities, the bound on $j$ \bound\ and $\Delta_V\geq 
{r_V \over 2}$.

Combining now \chitac\ and \phystac, we obtain $j$ in terms of $\Delta_V$:
\eqn\jdel{j+1={k\over 2}(1-2\Delta_V)~.}
The bound $j>-\half$ imposes that $\Delta_V<\half \left(1-{1\over k}\right)
<\half$. Requiring now the GSO condition \gsotac, we get:
\eqn\delrv{\pm(1-2\Delta_V)-r_V=2n+1~,}
where $n$ is an integer. Choosing the $+$ sign, so that the spacetime operator
would be chiral, we get $\Delta_V=n'-{r_V\over 2}$, with $n'\geq 1$.
However $\Delta_V\geq {r_V\over 2}$ implies $r_V \leq n'$ so that
$\Delta_V\geq n'-{n'\over 2}={n'\over 2} \geq \half$, in contradiction
with $\Delta_V<\half$. Therefore starting with an operator $V$ with
positive $R$-charge it is impossible to get a chiral spacetime operator.

Choosing thus the $-$ sign in \delrv, one gets $\Delta_V=n'' +{r_V\over 2}$,
with $n''\geq 0$. Clearly $\Delta_V<\half$ imposes $n''=0$ and thus
$V$ must be a worldsheet chiral operator. Moreover, its $R$-charge has
to satisfy the bound $r_V<1-{1\over k}$, which is listed in \chix.

The conclusion is that physical vertex operators like \nsvop\ only 
lead to anti-chiral operators in spacetime, with spacetime $R$-charge
given by $\sqrt{2k}q=-2(j+1)=-k(1-r_V)$. To obtain spacetime chiral operators
from vertex operators of the form \nsvop, we have to take an anti-chiral
operator $\tilde{V}$ of the $\NN/U(1)$ worldsheet CFT. It amounts to taking
the complex conjugate of \chix.

{}For vertex operators like \vgrav, where:
\eqn\psiphi{(\psi\Phi_j)_{j-1,m}=\psi^+ \Phi_{j,m-1}+\psi^- \Phi_{j,m+1}
-2\psi^3 \Phi_{j,m}~,}
the way one obtains the final result \chiw\ is very similar to the one
above. The presence of the $SL(2)$ fermions only modifies the GSO
and the on-shell conditions to \gsograv\ and \physgrav, 
and the spacetime weight. Otherwise the
reasoning is identical, giving (for $r_V\geq 0$) 
only spacetime chiral operators.

In the Ramond sector, we now show that the BRST condition is non-trivial
and actually fixes the spacetime chiral nature of \chiy.
We start from the vertex operators:
\eqn\rvop{\YY=e^{-{\varphi\over 2}}(S e^{iqY}V \Phi_j)_{j-\half,m}~,}
where the presence of a general spin field of the form \spigen\ affects
the total $Y$ and $Z$ charges, and we have taken the $SL(2)$ combination:
\eqn\sphi{(S\Phi_j)_{j-\half,m}= S_\half \Phi_{j,m-\half}-
S_{-\half}\Phi_{j,m+\half}~.}
Rewriting the spin field $S_r^{\epsilon_2\epsilon_Z}=s_r^{\epsilon_2}
e^{i\epsilon_Z({a\over 2}Z-{1\over \sqrt{2k} }Y)}$,
we can expand \rvop\ as:
\eqn\rvopex{\YY=e^{-{\varphi\over 2}}e^{i(q-{\epsilon_Z \over
\sqrt{2k} })Y} v e^{i({r_V \over a}+\epsilon_Z {a\over 2})Z}
(s \Phi_j)_{j-\half,m}~.}
Its on-shell condition is:
\eqn\onshr{{3\over 8}+\half \left(q-{\epsilon_Z \over \sqrt{2k} }
\right)^2 +\Delta_V-{r_V^2\over 2a^2} +\half \left(
{r_V \over a}+\epsilon_Z {a\over 2}\right)^2 +{1\over 4}-{j(j+1)\over k}=1~.}
It can be rearranged as:
\eqn\onshrbis{-{1\over k}\left(j+\half\right)^2+
\half \left(q-{\epsilon_Z \over \sqrt{2k} }\right)^2 +\Delta_V
+\epsilon_Z {r_V\over 2}=0~.}
Requiring that \rvop\ be (anti-)chiral in spacetime, we get the condition:
\eqn\rchi{j+\half=\pm \sqrt{k\over 2}\left(q-{\epsilon_Z \over \sqrt{2k} }
\right)~, }
so that the on-shell condition \onshrbis\ boils down to 
$\Delta_V=-\epsilon_Z {r_V\over 2}$. Since we chose $r_V\geq 0$, we
see that we must take $\epsilon_Z=-1$ and require that $V$ be a chiral 
operator of $\NN/U(1)$.

We will now impose the BRST condition on \rvop, taking into account that
$V$ is chiral. This entails that $G_{\NN/U(1)}(z)V(z')\sim G^-(z)V(z')
\sim (z-z')^{-1} V'(z')$. On the other hand, since $\epsilon_Z=-1$,
we have that $G^-(z) e^{-i{a\over 2}Z}(z')\sim (z-z')^{1/2} W(z')$,
so that again $G_{\NN/U(1)}$ does not give terms relevant to the BRST
condition when acting on \rvop.

We can rewrite once more \rvop, taking now into account \sphi\ and \pref:
\eqn\rvopter{\YY=e^{-{\varphi\over 2}}e^{i(q+{1\over \sqrt{2k} })Y}
e^{-i{a\over 2}Z}V \left(e^{-{i\over 2}(H_1 +\epsilon_2 H_2)}
\Phi_{j,m-\half}+\epsilon_2 c_1 c_2 e^{{i\over 2}(H_1 +\epsilon_2 H_2)}
\Phi_{j,m+\half}\right)~, }
and take the OPE with $G_{rest}=G_{SL(2)}+G_{U(1)}$ as given in 
\bosocur. The result is:
\eqn\rvopbrst{G_{rest}(z)\YY(z')\sim {1\over \sqrt{k}}(z-z')^{-3/2}
\left(\epsilon_2 (j+1) +{1-\epsilon_2\over 2}+\sqrt{k\over 2}q\right)
\YY'(z')~,}
where:
\eqn\yyprime{\YY'=e^{-{\varphi\over 2}}e^{i(q+{1\over \sqrt{2k} })Y}
e^{-i{a\over 2}Z}V \left(c_2 e^{-{i\over 2}(H_1 -\epsilon_2 H_2)}
\Phi_{j,m-\half}-\epsilon_2 c_1 e^{{i\over 2}(H_1 -\epsilon_2 H_2)}
\Phi_{j,m+\half}\right)~.}
Thus for $\epsilon_2=-1$ the BRST condition becomes $j=\sqrt{k\over 2}q$,
that is the vertex operator \rvop\ is spacetime chiral, while
for $\epsilon_2=1$ the BRST condition is $j+1=-\sqrt{k\over 2}q$,
so that the vertex operator \rvop\ is spacetime anti-chiral.

Taking $\epsilon_2=-1$, which is the choice in \vram--\spif, we now
show that the GSO projection fixes the spacetime
$R$-charge. First of all mutual locality of \vram--\spif\
with the supercharges \gspace--\splus\ is:
\eqn\mutloc{-{1\over 4}+rs\mp rs\pm {a^2\over 4} \mp {r_V\over 2}
\pm {1\over 2k} \pm {q\over \sqrt{2k}}\in {\bf Z}~,}
which translates to $\sqrt{2\over k}q-r_V \in 2{\bf Z}+1$, as in \gsoram.
This leads to:
\eqn\jrn{{2j\over k}-r_V=2n+1~, \qquad \qquad n\in {\bf Z}~.}
Taking into account the bound on $j$ \bound\ and that $0\leq r_V \leq 
3-{2\over k}={c_{\NN/U(1)}\over 3}$, we get that $n$ has to satisfy
the bound $-2+{1\over 2k}<n<-{1\over 2k}$ which implies $n=-1$.
We thus finally obtain $j={k\over 2}(r_V-1)$, and the physical
operators \chiy\ have spacetime $R$-charge $\sqrt{2k}q+1=2j+1=
1+k(r_V-1)$ (provided $r_V$ satisfies the bound \rbound).

{}Focusing now on operators which are spacetime anti-chirals, we take 
$\epsilon_2=1$ in \rvopter. The GSO condition changes slightly,
i.e. the third term of \mutloc\ has the opposite sign. This leads
to the condition $\sqrt{2\over k}q-r_V \in 2{\bf Z}$. Repeating a
very similar reasoning as above, we obtain the operators \chiyt\ with
$j+1={k\over 2}(2-r_V)$
and a spacetime $R$-charge of $\sqrt{2k}q+1=1+k(r_V-2)$.
It is shown in section 4 that these operators do not define a new
family of operators, but are actually complex conjugates of already
existing operators of the kind \chiy.

\subsec{Twisted Sector Physical Vertex Operators}
In section 4.2 we learn that spacetime (anti-)chiral operators coming from
the worldsheet twisted sectors are obtained twisting physical vertex
operators in the untwisted sector, which lead to (anti-)chiral operators
in spacetime.
Here we show in more detail how the on-shell condition of the twisted
vertex operators is enforced.

Starting with the operators \chix\ and twisting them with $t^w_\pm$,
we obtain operators:
\eqn\twix{e^{-\varphi}e^{i\sqrt{2\over k}(-j-1\pm {k\over 2}w)Y}V
\Phi^w_{jm}~,}
which are on-shell provided:
\eqn\twionsh{{1\over k}\left(-j-1\pm {k\over 2}w\right)^2 +{r_V\over 2}
-{j(j+1)\over k}-mw -{k w^2\over 4}=\half~,}
where we have used \sdimphiw. Note that the GSO condition is unchanged
since the twist fields are mutually local with the supercharges by
construction.
Using the relation between $j$ and $r_V$ in \chix, \twionsh\ boils down to:
\eqn\mj{m=\mp(j+1)~.}
Note that the spacetime weight is $h=-m-{k\over 2}w=\pm(j+1)-{k\over 2}w$
(recall that $w<0$ for ingoing states)
and the $R$-charge is ${R\over 2}=-(j+1)\pm {k\over 2}w=\mp h$, so that
twisting an operator \chix\ (which is anti-chiral in spacetime) with
$t_+^w$ leads to an anti-chiral operator, while twisting it with $t_-^w$
leads to a chiral operator in spacetime. Moreover \mj\ also tells us
that the original $\Phi_{jm}$ must be built respectively from a $\DD_j^-$
and a $\DD_j^+$ representation of $SL(2)$. In section 4.2 it is shown
that the operators above twisted by $t_-^w$ can be reconducted to 
the twist of operators of the kind \chiw.

Twisting the operators \chiw\ we obtain:
\eqn\twiw{e^{-\varphi}e^{i\sqrt{2\over k}(j\pm {k\over 2}w)Y}V
(\psi\Phi^w_j)_{j-1,m}~,}
where $(\psi\Phi^w_j)_{j-1,m}$ is given by \psiphi\ with $\Phi_{jm}$
replaced by $\Phi^w_{jm}$. The GSO projection is again unchanged,
however the on-shell condition is slightly complicated by the 
fact that $\Phi^w_{jm}\propto e^{iwH_1}e^{\sqrt{2\over k+2}(m+{k+2\over 2}w)x}$
(see \twt) and thus depends on $H_1$. Using also $\psi^\pm=e^{\mp iH_1}$,
we obtain that:
\eqn\delpsiphi{\Delta[(\psi\Phi^w_j)_{j-1,m}]=
\half -{j(j+1)\over k}-mw -{kw^2\over 4}~.}
Then the on-shell condition is:
\eqn\onshw{{1\over k}\left(j\pm {k\over 2}w\right)^2+{r_V\over 2}
-{j(j+1)\over k}-mw -{kw^2\over 4} =0~,}
and leads to $m=\pm j$. This value is allowed since 
$(\psi\Phi^w_j)_{j-1,m=\pm j}=\psi^\mp \Phi^w_{j, \pm(j+1)}$, while
the other terms in \psiphi\ annihilate the vacuum.
In this case too twisting with $t^w_+$ or $t^w_-$ leads respectively 
to anti-chiral and chiral operators in spacetime. The identifications
between twisted operators of the different kinds is listed in \alltwi.

Twisting now the Ramond sector operators \chiy, we obtain:
\eqn\twiy{e^{-{\varphi \over 2}} e^{i\sqrt{2\over k}(j+\half\pm {k\over 2}w)Y}
e^{-i{a\over 2}Z}V (s\Phi^w_j)_{j-\half, m}~,}
where:
\eqn\twiybis{\eqalign{(s\Phi^w_j)_{j-\half, m}=&
e^{-i(\half-w)H_1 +{i\over 2}H_2}\Psi_{j,m-\half}e^{i\sqrt{2\over k+2}
(m-\half +{k\over 2}w)x}\cr & -c_1 c_2 e^{i(\half+w)H_1 -{i\over 2}H_2}
\Psi_{j,m+\half}e^{i\sqrt{2\over k+2} (m+\half +{k\over 2}w)x}~.}}
The on-shell condition leads to:
\eqn\mjram{m=\pm\left(j+\half\right)~.}
The resulting operators are still BRST invariant, as one can check
with \bosocur, using \opes\ and discarding all operators which
annihilate the vacuum. Again the twisted operators are identified
in \alltwi, and discussed in section 4.2.

We have thus concluded our list of physical vertex operators leading
to (anti-)chiral operators in spacetime.

\appendix{B}{Cyclic Orbifolds}
In this appendix we collect some results about cyclic orbifolds
which are relevant to the discussion of CFTs on symmetric products in 
section 5.

Assume that we  take a general CFT $\MM$  with 
some symmetry algebra (which for convenience we will consider to be of rank 1
and designate its Cartan generator by $J$), then the generalized partition
function is defined to be:
\eqn\prtdef{Z(\tau,z,\bar{\tau},\bar{z})=q^{-{c\over24}} 
{\bar{q}}^{-{\bar{c}\over24}} {\rm Tr} q^{L_0} {\bar{q}}^{\bar{L}_0} x^{J}
{\bar{x}}^{{\bar{J}}}=\sum_{k,l} N_{kl} \chi_{k}(\tau,z) \bar{\chi}_{k}
(\bar{\tau},\bar{z})~. }
The generalized characters are defined as:
\eqn\genchar{{\chi}_{k}(\tau,z)={\rm Tr}_{\HH_{k}}e^{2\pi izJ}
e^{2\pi i \tau (L_0 -{c\over24})}~. }
$\chi_k$ is the character corresponding to a primary field $\psi_k$
where the trace is taken on the Virasoro algebra representation space $\HH_k$
(i.e. the descendants) and $q=e^{2\pi i \tau}$, $x=e^{2\pi iz}$.
We now look at the cyclic orbifold
$\MM^N/Z_N$ for $N$ prime. In \refs{\klsc,\fks} 
the modular invariant partition function of such a cyclic orbifold model is
shown to be:
\eqn\mipfco{\ZZ(\tau,z)={Z^{N}(\tau,z)\over N}+{(N-1)\over N} 
\left[ Z(N\tau,Nz) + \sum_{n=0}^{N-1} Z({\tau+n \over N},z) \right]~.}
This is obtained by taking the $Z_N$ invariant projection on the $\MM^N$
partition function and then adding the terms required by modular invariance.
The first two terms correspond to the untwisted sector, and the last ones 
to the twisted sectors. 
Note that the $N-1$ factor in front of the twisted part appears
because we have $N-1$ identical twisted sectors in a $Z_N$ orbifold (when
$N$ is prime).
{}From this very general formula one can already extract the vacuum energies 
and $J$ charges of states in the twisted sectors \refs{\klsc,\fks}. 
To compute the vacuum energies in the twisted sectors note
that the modular parameter in the twisted part of the partition function
has become ${\tau+n \over N}$. Plugging this into the definition of the 
character and rearranging to get an expression of the form 
\eqn\char{\chi(\tau)=q^{-{\tilde{c}\over24}} {\rm Tr}_{\HH} 
q^{\tilde{L}_0}~, } 
where $\tilde{c}=Nc$ and $\tilde{L}_0$ are the central
charge and energy of the twisted theory in the appropriate twisted sector, 
we get:
\eqn\twistchar{\eqalign{
{\chi}_{k}({\tau+n \over N})&=e^{2 \pi i({\tau+n \over N})
(-{c\over24})}{\rm Tr}e^{2 \pi i({\tau+n \over N})L_0}=q^{-{c\over24N}}  
{\rm Tr}q^{{L_0 \over N}}    
e^{2 \pi i({n \over N})(L_0-{c\over24})}\cr
&= q^{-{Nc\over24}} {\rm Tr} q^{ {L_0 \over N}
+{c\over24}( N-{1\over N} )} e^{2 \pi i({n \over N})(L_0-{c\over24})}~.\cr}}
The term ${1\over{N}} \sum_{n=0}^{N-1}{\rm exp}(2 \pi i({n \over {N}})
(L_0-{c\over24})) $ together with the anti-holomorphic contribution
serves as a projection operator onto $h-\bar{h} =0 \ ({\rm mod} \ N)$. 
We can thus identify the weight of states
generated by an operator of the original theory (with weight $h$) in the 
$Z_{N}$ twisted sectors of the orbifold to be:
\eqn\hnew{h^N={h+m \over{N}}+{c\over24}{N^2-1\over N}~, }
where $m=0,1,\dots, N-1$ and 
we denote by $N$ quantities in the twisted sectors.
This means that every primary field and its first $N-1$ descendants in the 
original theory  $\MM$  give rise to a primary field in the $Z_{N}$ orbifold
theory with weight given by \hnew.
In particular this gives the vacuum energies in the twisted sectors 
as ${c\over24}{N^2-1\over N}$. 
Every state of a given weight in the twisted sector has multiplicity $N-1$.

Another conclusion arises from examining the second argument of the twisted 
part of the orbifold partition function in \mipfco \ which is $Z({\tau+n
\over N},z)$. This shows that the symmetry charges ($J$ in our case) 
carried by the operators in the $Z_{N}$ twisted
sectors are identical to those in the untwisted sector \fks, we thus write:
\eqn\qnew{R^N=R~.}
We now generalize the expressions for non-prime $N$.
This is derived in essentially the same way but now we should remember
that there are elements of the group $Z_{N}$ that have degree $r\leq N$ and so
complete a full cycle in $r$ steps, i.e. $b^r=1$. 
In the above derivation the fact that $N$ was prime 
was used to make sure that all the elements are of degree $N$ and thus
all twisted sectors are equivalent.
{}For non-prime $N$ there are elements in $Z_{N}$ whose degree $r$ is a divisor
of $N$. The sectors twisted by these elements and their powers will
lead to terms in the partition function which are similar to the ones
in \mipfco\ but with $N$ replaced by $r$, and each one to the power of $N/r$
since this is the number of disconnected cycles.
Therefore we should change \mipfco \ to:
\eqn\mipfconp{\ZZ(\tau,z)={Z^N(\tau,z)\over N}+{1\over N}
\sum_{m=1}^{N-1}\left[ 
Z^{{N\over r_{m}}}(r_{m}\tau,r_{m}z) +
\sum_{n=0}^{r_{m}-1} Z^{{N\over r_{m}}}({\tau+n \over r_{m} },z) 
\right] + \dots~,}
where $r_m$ is the degree of $a^m$, with $a$ a generator of $Z_N$ ($r_1=N$). 
We see that for $N$ prime the degree of all the elements $r_{m}=N$ is 
independent of $m$ and we recover \mipfco .
The terms omitted from this formula 
describe sectors where the twist in the timelike direction of the torus
corresponds to
an element of $Z_N$ with higher degree then the element corresponding to 
the twist in the spacelike direction.
The outcome of this discussion is that for any $N$, 
there are twisted states obeying 
\hnew\ and \qnew, however their multiplicity is less than $N-1$ if $N$ is not
prime.

Note that although in the $\MM^N/Z_N$ orbifold the $Z_N$ twist fields come with
non-trivial multiplicities, when one considers the full $\MM^p/S_p$
orbifold these same twist fields actually come with multiplicity one.
The reason is that in the $S_p$ group, a $Z_N$ cycle is one conjugacy
class and thus gives rise to only one twisted sector.

The explicit forms of the $Z_N$ partition function \mipfco\ and \mipfconp\
allows us to heuristically distinguish between the single particle
and multi-particle spectrum, by simply noting the power to which 
the partition function of $\MM$ is raised. The terms
related to twisted sectors of maximal degree $N$ appear to unit power
and thus correspond to single particle states, while for instance
the first term $Z^N$ indicates that in the untwisted sector all the states 
that are not in the diagonal $\MM$ arise from multiplying operators
in the various copies, and thus are identified with multi-particle states.

\listrefs

\end